\documentclass[11pt]{article}
\usepackage{amsmath,amsthm,latexsym,amssymb,amsfonts,epsfig}

\makeatletter
\@addtoreset{equation}{section}
\makeatother

\newtheorem{Definition}{Definition}[section]

\def\be{\begin{equation}}
\def\ee{\end{equation}}
\def\ba{\begin{eqnarray}}
\def\ea{\end{eqnarray}}
\def\Nl{{\mathchoice
{\setbox0=\hbox{$\displaystyle\rm N$}\hbox{\hbox to0pt
{\kern0.4\wd0\vrule height0.9\ht0\hss}\box0}}
{\setbox0=\hbox{$\textstyle\rm N$}\hbox{\hbox to0pt
{\kern0.4\wd0\vrule height0.9\ht0\hss}\box0}}
{\setbox0=\hbox{$\scriptstyle\rm N$}\hbox{\hbox to0pt
{\kern0.4\wd0\vrule height0.9\ht0\hss}\box0}}
{\setbox0=\hbox{$\scriptscriptstyle\rm N$}\hbox{\hbox to0pt
{\kern0.4\wd0\vrule height0.9\ht0\hss}\box0}}}}
\def\Zl{{\mathchoice
{\setbox0=\hbox{$\displaystyle\rm Z$}\hbox{\hbox to0pt
{\kern0.4\wd0\vrule height0.9\ht0\hss}\box0}}
{\setbox0=\hbox{$\textstyle\rm Z$}\hbox{\hbox to0pt
{\kern0.4\wd0\vrule height0.9\ht0\hss}\box0}}
{\setbox0=\hbox{$\scriptstyle\rm Z$}\hbox{\hbox to0pt
{\kern0.4\wd0\vrule height0.9\ht0\hss}\box0}}
{\setbox0=\hbox{$\scriptscriptstyle\rm Z$}\hbox{\hbox to0pt
{\kern0.4\wd0\vrule height0.9\ht0\hss}\box0}}}}
\def\Ql{{\mathchoice
{\setbox0=\hbox{$\displaystyle\rm Q$}\hbox{\hbox to0pt
{\kern0.4\wd0\vrule height0.9\ht0\hss}\box0}}
{\setbox0=\hbox{$\textstyle\rm Q$}\hbox{\hbox to0pt
{\kern0.4\wd0\vrule height0.9\ht0\hss}\box0}}
{\setbox0=\hbox{$\scriptstyle\rm Q$}\hbox{\hbox to0pt
{\kern0.4\wd0\vrule height0.9\ht0\hss}\box0}}
{\setbox0=\hbox{$\scriptscriptstyle\rm Q$}\hbox{\hbox to0pt
{\kern0.4\wd0\vrule height0.9\ht0\hss}\box0}}}}
\def\Rl{{\mathchoice
{\setbox0=\hbox{$\displaystyle\rm R$}\hbox{\hbox to0pt
{\kern0.4\wd0\vrule height0.9\ht0\hss}\box0}}
{\setbox0=\hbox{$\textstyle\rm R$}\hbox{\hbox to0pt
{\kern0.4\wd0\vrule height0.9\ht0\hss}\box0}}
{\setbox0=\hbox{$\scriptstyle\rm R$}\hbox{\hbox to0pt
{\kern0.4\wd0\vrule height0.9\ht0\hss}\box0}}
{\setbox0=\hbox{$\scriptscriptstyle\rm R$}\hbox{\hbox to0pt
{\kern0.4\wd0\vrule height0.9\ht0\hss}\box0}}}}
\def\Cl{{\mathchoice
{\setbox0=\hbox{$\displaystyle\rm C$}\hbox{\hbox to0pt
{\kern0.4\wd0\vrule height0.9\ht0\hss}\box0}}
{\setbox0=\hbox{$\textstyle\rm C$}\hbox{\hbox to0pt
{\kern0.4\wd0\vrule height0.9\ht0\hss}\box0}}
{\setbox0=\hbox{$\scriptstyle\rm C$}\hbox{\hbox to0pt
{\kern0.4\wd0\vrule height0.9\ht0\hss}\box0}}
{\setbox0=\hbox{$\scriptscriptstyle\rm C$}\hbox{\hbox to0pt
{\kern0.4\wd0\vrule height0.9\ht0\hss}\box0}}}}
\def\Hl{{\mathchoice
{\setbox0=\hbox{$\displaystyle\rm H$}\hbox{\hbox to0pt
{\kern0.4\wd0\vrule height0.9\ht0\hss}\box0}}
{\setbox0=\hbox{$\textstyle\rm H$}\hbox{\hbox to0pt
{\kern0.4\wd0\vrule height0.9\ht0\hss}\box0}}
{\setbox0=\hbox{$\scriptstyle\rm H$}\hbox{\hbox to0pt
{\kern0.4\wd0\vrule height0.9\ht0\hss}\box0}}
{\setbox0=\hbox{$\scriptscriptstyle\rm H$}\hbox{\hbox to0pt
{\kern0.4\wd0\vrule height0.9\ht0\hss}\box0}}}}
\def\Ol{{\mathchoice
{\setbox0=\hbox{$\displaystyle\rm O$}\hbox{\hbox to0pt
{\kern0.4\wd0\vrule height0.9\ht0\hss}\box0}}
{\setbox0=\hbox{$\textstyle\rm O$}\hbox{\hbox to0pt
{\kern0.4\wd0\vrule height0.9\ht0\hss}\box0}}
{\setbox0=\hbox{$\scriptstyle\rm O$}\hbox{\hbox to0pt
{\kern0.4\wd0\vrule height0.9\ht0\hss}\box0}}
{\setbox0=\hbox{$\scriptscriptstyle\rm O$}\hbox{\hbox to0pt
{\kern0.4\wd0\vrule height0.9\ht0\hss}\box0}}}}

\title{{\sf Semiclassical analysis}\\ 
{\sf of the}\\ 
{\sf Loop Quantum Gravity volume operator:}\\
{\sf I. Flux Coherent States} 
}

\author{{\sf C. Flori$^1$}\thanks{{\sf cecilia.flori@aei.mpg.de}}, 
{\sf T.
Thiemann$^{1,2}$}\thanks{{\sf
thiemann@aei.mpg.de,tthiemann@perimeterinstitute.ca}} \\
\\
{\sf $^1$ MPI f. Gravitationsphysik, Albert-Einstein-Institut,} \\
{\sf Am M\"uhlenberg 1, 14476 Potsdam, Germany}\\
\\
{\sf $^2$ Perimeter Institute for Theoretical Physics,}\\
{\sf 31 Caroline Street N, Waterloo, ON N2L 2Y5, Canada}}
\date{{}}

\begin{document}

\maketitle

\begin{abstract}
The volume operator plays a pivotal role for the quantum dynamics of 
Loop 
Quantum Gravity (LQG), both in the full theory and in truncated models 
adapted to cosmological situations coined Loop Quantum Cosmology (LQC). 
It is therefore crucial to check whether its semiclassical limit 
coincides with the classical volume operator plus quantum corrections.

In the present article we investigate this question by
generalizing and employing previously 
defined coherent states for LQG which derive from a 
cylindrically consistently defined complexifier operator which is the 
quantization of a known classical function. 
These coherent states are not normalizable due to the non separability 
of the LQG Hilbert space but they define uniquely define cut -- off 
states depending on a finite graph. 

The result of our analysis is that the 
expectation value of the volume operator with respect to coherent states 
depending on a graph with only $n-$valent verticies reproduces its 
classical value at the phase space point at which the coherent state is 
peaked only if $n=6$. In other words, the semiclassical sector of LQG 
defined by those states is described by graphs with cubic topology!
This has some bearing on current spin foam models which are all based 
on 
four valent boundary spin networks.
\end{abstract}

\newpage

\tableofcontents

\newpage

\section{Introduction}
\label{s1}

The volume operator for Loop Quantum Gravity\footnote{See \cite{1a,1b} 
for 
recent books and \cite{2} for recent reviews.} (LQG) enters the quantum 
dynamics of the theory in a prominent way. Without it, Hamiltonian 
constraint operators \cite{3}, Master constraint operators \cite{4} or 
physical Hamiltonian operators \cite{5} cannot be defined. This is also 
true for truncated models of LQG such as Loop Quantum Cosmology (LQC)
\cite{6} which is supposed to describe well the homogeneous sector of the
theory. In LQC one generically finds that the quantum evolution 
of operators corresponding to classically singular Dirac 
observables remain finite. This feature can be traced back to the 
way that inverse powers of the local volume, which enter into the 
expression for the triad, cotriad and other types couplings between 
geometry and geometry or geometry and matter, are quantized in the full 
theory \cite{3}. Namely such derived operators are obtained as commutators 
between holonomy operators and powers of the volume operator.

In view of its importance, it is of outmost interest to verify that the 
classical limit of the LQG volume operator coincides with the classical
volume. By this we mean that the expectation value of the volume operator 
with respect to suitable semiclassical states which are peaked on a 
given point in phase space coincides with the value of the classical 
volume at that phase space point up to small corrections and that its 
fluctuations are small.

Notice that there are actually two versions of the volume operator
discussed in the literature \cite{7,8} which result from nonequivalent 
regularisations of the products of operator valued distributions that 
appear at intermediate stages. However, only the operator in \cite{8}
survives the consistency test of \cite{9}, namely that writing 
the volume in terms of triads which then are quantized using the 
above mentioned commutator, delivers the same operator up to $\hbar$ 
corrections as the direct quantization. This check is not unimportant as 
otherwise we should not trust the triad and cotriad quantisations that 
enter the quantum dynamics.

The semiclassical analysis of the volume operator has not been carried out 
yet although in principle suitable semiclassical (even coherent) states 
for LQG are available \cite{10}. The reason for this is that the spectral
decomposition (projection valued measure) of the volume operator cannot 
be computed exactly in closed form which however is needed in order 
to do exact practical calculations. More precisely, the volume operator
is the fourth root of a positive operator $Q$ whose matrix elements can be 
computed in closed form \cite{11} but which cannot be diagonalized 
analytically. More in detail, the volume operator has discrete (that is, 
pure point) spectrum and it attains a block diagonal form where the blocks 
are labelled by the graphs and spin quantum numbers (labelling the edges 
of the graph) of spin network functions (SNWF) \cite{12}.
SNWF's form a convenient
basis of the LQG Hilbert space \cite{13} which is the unique 
(cyclic) Hilbert which carries a unitary representation of the
spatial diffeomorphism group and the Poisson$^\ast-$algebra of 
the elementary (flux and holonomy) variables \cite{14}. 
The blocks turn out to be finite dimensional matrices whose matrix 
elements can be 
expressed in terms of polynomials of 6j symbols which fortunately can be
avoided\footnote{Notice that Racah's formula provides a closed expression 
for the 6j symbol but that it involves implicit sums and factorials
involving large integers which quickly becomes unmanageable even 
numerically.} using a telescopic summation technique \cite{15} related 
to 
the Elliot -- Biedenharn identity \cite{16} so that a manageable closed 
expression results. However, the size of these matrices grows 
exponentially with increasing spin quantum numbers and since the
expression for coherent states is a coherent superposition of SNWF's with 
arbitrarily large spin quantum numbers, a numerical computation of 
the expectation value using 
numerical diagonalisation techniques which are currently being developed 
\cite{17} is so far out of reach\footnote{The coherent superposition 
contains a damping factor which suppresses large spins and thus large 
matrices so that one may truncate the involved infinite series over spin 
quantum numbers at finite values making only negligible errors, but still
the computational effort is currently too high for supercomputers, see 
e.g. the computation time estimates in reported \cite{17}.} 

The way to make progress is to use semiclassical perturbation theory
developed in \cite{18} and already applied in \cite{19,20}. The idea is 
quite 
simple: In practical calculations one needs the expectation value 
of $Q^q$ where $q$ is a rational number in the range $0<q\le \frac{1}{4}$.
Introduce the ``perturbation operator'' $X:=\frac{Q}{<Q>}-1$ where 
the expectation value $<Q>$ of the positive operator $Q$ is exactly 
computable. Notice that $X$ is bounded from below by $-1$. Then trivially 
$<Q^q>=<Q>^q\;<[1+X]^q>$. Now we use that there exist positive numbers 
$p$ such that the classical estimates $1+qx-px^2\le
(1+x)^q\le 1+q x$ are valid for all $x\ge -1$. Finer estimates of this 
form involving arbitrary powers of $x$ are also available \cite{19}. In 
view of the 
spectral theorem, this classical estimate survives at the quantum level
and we have $Y_-\le Y\le Y_+$ where $Y_+=1+qX,\;Y_-=Y_+ -p X^2,\;
Y=(1+X)^q$. It follows that $<Y>\in [<Y_+>-p<X^2>,<Y_+>]$. However,
$<Y_+>=1$ and $<X^2>=[<Q^2>-<Q>^2]/<Q>^2$ is proportional to the 
relative fluctuation of $Q$ which of order of $\hbar$ \cite{10}. 
It follows that to zeroth order in $\hbar$ we may replace $<Q^q>$
by $<Q>^q$ which is computable and whose error estimates are also 
computable as shown above.

While possible, the exact computation of $<Q>,\;<Q^2>,\;..$ is still quite
involved. Here the following result, proved in \cite{10}, is convenient:
The computation of these expectation values (more generally for low 
order polynomials in flux operators) for 
$SU(2)$ spin 
network states coincide, to zeroth order in $\hbar$, with the 
corresponding calculations for $U(1)^3$ spin networks. On those, the 
volume operator is even diagonal. Hence we conclude that, as long as we 
are only interested in the zeroth order in $\hbar$ contribution, we may 
evaluate the expectation value of the volume operator for a fictive theory
in which we may replace the non Abelian group $SU(2)$ by the Abelian group 
$(U1)^3$ which makes all calculations dramatically simpler.

The coherent states developed for LQG so far all have been constructed 
using the complexifier method reviewed in \cite{21} which generalizes 
the coherent state construction for phase spaces that are cotangent 
bundles over compact groups developed in \cite{22}\footnote{See also 
\cite{23}
for related ideas valid for Abelian gauge theories such as Maxwell theory 
or linearized gravity.}.
This involves the heat kernel evolution of the delta distribution
(which is the matrix element of the unit operator in the Schr\"odinger
(position) representation) with respect to a generalized Laplace 
operator, called the complexifier, followed by a certain analytic 
continuation.
Now the unit operator in LQG can be written as a resolution of unity in 
terms of SNWF's and although the heat kernel is a damping operator,
since the SNWF are not countable (the LQG Hilbert space is not separable),
the resulting expression is not normalizable. It defines a well defined 
distribution (in the algebraic dual of the finite linear span of SNWF's)
which can be conveniently written as a sum over cut -- off states 
labelled by finite graphs. These states, called ``shadows'' in 
\cite{22}, {\it are} normalizable and one can use those in order to 
perform semiclassical calculations. Of course, one expects that only cut 
-- off states labelled by graphs which are sufficiently fine with respect 
to the classical three metric to be approximated are good semiclassical 
states. 

Hence we see that the input in the semiclassical calculations consists in 
the choice of a complexifier and the choice of a graph. One may wonder 
why the complexifier has to be chosen and is not dictated by the dynamics 
of the theory, that is, that the coherent states remain coherent under 
quantum time evolution (e.g. for the harmonic oscillator, the complexifier  
is strictly the Laplacian on the real line). The reason is twofold:\\
On the one hand one could perform constraint quantization and then  
we are working at the level of the kinematical 
Hilbert space on which the quantum constraints have not been imposed yet
and all we want to make sure is that the volume operator, which is not 
gauge invariant (and thus does not preserve the physical Hilbert space) 
has a good classical limit on the kinematical Hilbert space (we would 
not trust a constraint quantization for which not even that was true).\\
On the other hand, one could also work at the level of the physical 
Hilbert space as outlined 
in \cite{5} and then one would expect that the time evolution of any 
reasonable choice of complexifier coherent states with respect 
to the physical Hamiltonian defined there remains coherent (and peaked on 
the classical trajectory) for sufficiently short time intervals. Notice 
that for interacting theories as simple as the an-harmonic oscillator 
globally stable coherent states have so far not been found. 

The choice of the complexifier will be guided by practical considerations, 
namely that it be diagonal on SNWF's and that it is a damping operator
which makes the heat kernel evolution of the delta distribution restricted 
to a graph normalizable. Moreover, it should be gauge invariant under the 
$SU(2)$ gauge group. As far as the choice of the graph is concerned, for 
practical reasons one will choose graphs that are topologically regular, 
that is, have constant valence for each vertex. Indeed, the semiclassical 
calculations performed in \cite{19,20} were done using graphs of cubic 
topology with good semiclassical results.

The existing literature on such coherent states for LQG can be divided 
into two classes, depending on a certain structure that defines them:\\
On the one hand, there are {\it gauge covariant flux} coherent states
which depend on collections of surfaces and path systems inside them
\cite{10}. On the other hand, there are {\it area} coherent states which 
only depend on collections of surfaces but involve area operators rather 
than flux operators \cite{21}. We will review and generalize both 
constructions.
In its most studied incarnations, the collection of surfaces involved 
in \cite{10} is defined by a polyhedronal partition of the spatial 
manifold while the collection involved in \cite{21} is in terms of a 
parquette of foliations of the spatial manifold. 

The novel results of this article are as follows:
\begin{itemize}
\item[1.] {\it Cylindrically consistent complexifier}\\
The important thing to 
notice is that both constructions are cylindrically consistent: Both 
derive from a complexifier which is graph independently defined in 
terms of the surfaces (and path systems) only and has 
consistent cylindrical projections to spin network functions over 
any graph. We stress this, because from the discussion in \cite{10} one 
may get the impression that the states constructed there do not come 
from a single complexifier operator. We clarify this in this paper by 
explicitly constructing those projections. The apparent contradiction
is resolved by observing that in \cite{10} cut -- off states were
displayed only for graphs which are dual to a fixed polyhedronal 
complex. In this paper we treat the general case which is more 
complicated. We also give a compact expression for the eigenvalues of 
the complexifier of type \cite{21} which is missing there. 
\item[2.] {\it Graphs of arbitrary valence}\\
The real motivation for the present article is that the calculations 
performed in \cite{19,20} raised the suspicion that the volume operator 
attains acceptable coherent state expectation values only for graphs 
of cubic topology. Therefore in this paper we perform the semiclassical 
analysis with respect to topologically regular graphs of valence $n$ with 
$n=4,6,8$. Ideally one would like to treat the general case but this 
leads to complicated book keeping problems. Fortunately, the cases
we consider turn out to be sufficient to answer the afore posed
question.
\end{itemize}
The outcome of our analysis is that, no matter whether one uses the 
states \cite{10} or \cite{21}, {\it the 
correct semiclassical limit is 
attained with these states for $n=6$ only}. In other words:\\
{\it Up to now, there are no semiclassical 
states known for other than cubic graph topology!}

We interpret this result as saying that the states we constructed 
are simply not semiclassical for the volume operator unless $n=6$.   
The ratio consisting of expectation value 
of the volume operator with respect to the coherent states for graphs of 
valence $n\not=6$ divided by the classical value deviates from unity 
by a geometrical factor $q_n$ depending on $n$ which is of order unity 
and not at all of order $\hbar$. One might argue that therefore one 
should rescale the volume operator by $1/q_n$. However, first of all 
this is not allowed because the normalization is fixed \cite{9} and 
secondly even if one would rescale, the statement would still be that 
the correct expectation value is attained for a unique valence only.

It may be that there are other semiclassical states coming from a 
different complexifier or from an entirely different method which does 
not have this problem. However, notice that all coherent states that 
were constructed so far for free field theories on Minkowski space come 
from a complexifier, so the complexifier method is natural and well 
tested in other contexts \cite{21}. Secondly, as shown in \cite{21}, 
the choice of a SU(2) invariant and background metric 
invariant complexifier for LQG with 
computable spectrum and the required damping behavior leaves relatively 
little possibilities and to the best of our knowledge, the states proposed 
in \cite{10,21} and used here are the simplest and only ones known so 
far. Therefore, even if better suited states exist, they will be very 
hard to guess and even harder to do practical calculations with. 

The implication of our result for LQG then seems to be that the 
semiclassical sector of the theory is spanned by SNWF based on cubic 
graphs. This has some bearing for spin foam models \cite{25} which are 
supposed to be -- but have so far not been proved to be -- the path 
integral formulation of LQG. Spin foams are 
certain state sum models  
based on simplicial triangulations of four manifolds whose dual graphs 
are therefore five valent. Since the intersection of this graph with a 
boundary three manifold is therefore four valent, we see that spin foam 
models based on simplicial triangulations corresponds to boundary Hilbert 
spaces spanned by spin network states based on four valent graphs 
only\footnote{
As an aside, whether this boundary Hilbert space of spin foams really 
can be 
interpreted as the four valent sector of LQG is a subject of current 
debate even with the recent improvement \cite{26} of the Barrett -- 
Crane model \cite{27} mostly studied so far. There are two problems:
First, the boundary connection predicted by spin foams does not coincide 
with the LQG connection \cite{28}. Secondly, the four valent sector 
of the LQG Hilbert space is not a superselection sector for the holonomy
flux algebra of LQG. In fact, the LQG representation is known not only 
to be a cyclic but even an irreducible representation \cite{28a}. 
Therefore the 
four valent sector is not invariant under the LQG algebra.}.

But even if these mismatches between LQG and spin foams could be 
surmounted, the result of our analysis seems to say that {\it the 
boundary 
Hilbert space of current spin foam models does not contain any 
semiclassical states!} This seems to contradict recent findings that 
the graviton propagator derived from spin foam models comes out 
correctly \cite{29}. However, notice that these results only show that 
the propagator comes out with the correct fall off behavior while 
the correct tensorial structure has not been verified yet. \\
An easy way to possibly repair this is to generalize spin foam models 
and
allow for arbitrary, in particular cubic, triangulations as suggested in 
\cite{30,30a}.  \\ 
\\
The outline of this paper is as follows:\\
\\

In section two we review the general complexifier method
and generalize the complexifier coherent states proposed in 
\cite{10,21}. In particular, we review the two families of coherent 
states introduced so far which are based on complexifiers 
constructed from squares of flux or area 
operators respectively.
We also explicitly compute the corresponding cut -- off 
states for SU(2) and show when and why it is justified to work with a 
fictive 
$U(1)^3$ theory instead as far as the zeroth order in $\hbar$ for the 
expectation values is concerned. 

In section three we construct explicitly convenient tetrahedronal, 
cubical and octaheronal cell decompositions of the spatial manifold 
which also define regular dual $n=4,6,8$ valent dual graphs. These 
constructions are needed for the explicit calculations in section four
and for our companion paper \cite{30b}. 

In section four we compute the expectation value of the volume 
operator for the constant valence cases $n=4,6,8$ for the states 
of type \cite{10}. We will do the same in our companion paper \cite{30b} 
for the states of type \cite{21}. In addition, in \cite{30b} we answer 
the following question: While the flux complexifier is built from a 
discrete set of fluxes, the area complexifier is built from a continuous 
set of areas whose underlying surfaces fill all of space. While this 
makes the calculations more cumbersome, it might improve the quality of 
the corresponding coherent states in the sense that the expectation 
values do not strongly depend on the cut -- off graph chosen (as already 
mentioned above, at least when the graph has cubic topology). Hence, we 
examine the question whether the expectation value of geometrical 
operators is invariant under Euclidean transformations of the graph
in case that the spatial metric to be approximated is Euclidean. 

In section five we summaries and conclude.

\section{Complexifier Coherent States}
\label{s2}

In this section we review the complexifier method to construct 
coherent states. We will be brief, all details can be found in \cite{21} 
and references therein. This section is divided into six parts:\\
In the first we define the general complexifier framework. In the second 
we provide natural gauge invariant choices of 
complexifiers for background independent SU(2) gauge theories 
such as General Relativity and fictive $U(1)^3$ gauge theories. 
In the third we compute the associated coherent states. These are not 
normalizable because they involve a sum over an uncountable set of 
graphs but one can extract from them normalizable states in which
the sum over all graphs is cut off to a finite number and which we 
display in the fourth part. In the fifth part we compare the properties 
of the states \cite{10} and \cite{21}. Finally, 
in the sixth we show that as far as the zeroth 
order in $\hbar$ is concerned, we may replace the complicated $SU(2)$ 
calculations by the much simpler $U(1)^3$ calculations under certain 
restrictions on the cut -- off graph. This approximation is needed only
for states of type \cite{21}. For the states of type \cite{10} it is 
sufficient to take as a cut -- off graph any graph which is dual to the 
polyhedronal complex that defines the complexifier.

\subsection{General Complexifier Method}
\label{s2.1}

We consider a symplectic manifold ${\cal M}=T^\ast({\cal C})$ which is a 
cotangent bundle over a configuration manifold $\cal C$ which may be 
infinite dimensional (we will suppress any indices in what follows). 
\begin{Definition} \label{def2.1} ~\\
A complexifier $C:\;{\cal M} \to \mathbb{R}^+$ is a sufficiently smooth, 
positive function on 
$\cal M$ with dimension of an action which has the following scaling 
behavior 
\be \label{2.1}
\lim_{\lambda \to \infty} \frac{C(q,\lambda p)}{\lambda}=\infty
\ee
where $(q,p)$ are the canonically conjugate, real configuration and 
momentum 
coordinates on $\cal M$.
\end{Definition} 
The reasons for these restrictions become evident in a moment. With the 
aid of $C$ we define 
\be \label{2.2}
z:=\exp(-i\{C,.\})\cdot q=\sum_{n=0}^\infty\;\frac{(-i)^n}{n!}\;
\{C,q\}_{(n)}
\ee
with inductively defined multiple Poisson brackets
$\{C,f\}_{(0)}:=f,\;\{C,f\}_{(n+1)}:=\{C,\{C,f\}_{(n)}\}$. The meaning 
of ``sufficiently smooth'' is that all coefficients in the Taylor 
expansion (\ref{2.2}) exist. Notice that (\ref{2.2}) defines a (complex 
valued) canonical transformation whence $\{z,z\}=\{\bar{z},\bar{z}\}=0$ 
(this is non trivial
when $\dim({\cal C})\ge 2$). The scaling behavior implies that 
$z,\;\bar{z}$ can be used as coordinates for $\cal M$. In fact, $z\in 
{\cal C}^{\mathbb{C}}$ defines a complex polarization of $\cal 
M$.  

We now assume that $\cal M$ can be quantized, that is, there is a 
representation $(q,p)\mapsto (\hat{q},\hat{p})$ of the 
Poisson$^\ast$ algebra defined by 
$\{q,q\}=\{p,p\}=0,\;\{p,q\}=1_{{\cal M}};\;\;
\bar{q}=q,\;\bar{p}=p$ on a Hilbert space of the form ${\cal 
H}=L_2(\overline{{\cal C}},d\mu)$ where $\overline{{\cal C}}={\cal C}$ 
in the finite dimensional case and otherwise ${\cal C}\subset 
\overline{{\cal C}}$ in the infinite dimensional case is a suitable 
distributional extension. That is, the operators satisfy (assuming 
careful domain definitions)
$[\hat{q},\hat{q}]=[\hat{p},\hat{p}]=0,\;[\hat{p},\hat{q}]=i\hbar\;1_{{\cal 
H}};\;\;
\hat{q}^\dagger=\hat{q},\;\hat{p}^\dagger=\hat{p}$. Here 
$\overline{{\cal C}}$ comes with some topology and $\mu$ is a Borel 
measure on it.

Assuming that also $C$ has a quantization $\hat{C}$ as a positive, self 
adjoint operator (in field theories this is non trivial due to 
operator ordering and operator product expansion questions) we construct 
the operator representation of (\ref{2.2}) 
by substituting Poisson brackets by commutators divided by $i\hbar$
\be \label{2.3}
\hat{z}:=\sum_{n=0}^\infty\;\frac{(-i)^n}{n!\;(i\hbar)^n}\;
\{\hat{C},\hat{q}\}_{(n)}=e^{-\hat{C}/\hbar}\;\hat{q}\;e^{\hat{C}/\hbar}
\ee
This formula explains the dimension restriction on $C$. The operator 
$e^{\pm \hat{C}/\hbar}$ is well defined via the spectral theorem.
We will refer to $e^{-\hat{C}/\hbar}$ as the {\it heat kernel} and to 
$\hat{z}$ as the {\it annihilation operator}. 

The $\delta$ distribution $\delta_{q_0}$ 
with support at $q_0\in 
\overline{{\cal C}}$ on the subset of ${\cal H}$ consisting of the 
continuous functions is defined by
\be \label{2.4}
\delta_{q_0}[\psi]:=\psi(q_0):=<\delta_{q_0},\psi>:=\int_{\overline{{\cal 
C}}}\;
d\mu(q)\;\delta_{q_0}(q) \psi(q)
\ee
where $\delta_{q_0}(q)$ is the integral kernel of the unit operator.
We define for $z\in \overline{{\cal C}}^{\mathbb{C}}$
the {\it coherent state}
\be \label{2.5}
\psi_z:=[e^{-\hat{C}/\hbar} \delta_{q_0}]_{q_0\to z}
\ee
It is defined as ``heat kernel evolution'' followed by analytic 
continuation. In order that this makes sense, the function 
$e^{-\hat{C}/\hbar} \delta_{q_0}$ must not only be in $\cal H$ but also 
analytic in $q_0$. This explains the positivity and scaling requirement 
on $C$ which 
makes sure that the heat kernel is a 
damping operator such that at least for separable $\cal H$ the function
(\ref{2.5}) is not only normalizable but also analytic. Here we have 
assumed that the map ${\cal C}\to {\cal C}^{\mathbb{C}};\;q\mapsto z$
in (\ref{2.2}) has an extension to some $\overline{{\cal 
C}}^{\mathbb{C}}$.

The whole point of this construction is that one can easily verify
\be \label{2.6}
\hat{z} \; \psi_z=z\; \psi_z
\ee
Thus, $\psi_z$ is an eigenfunction of the annihilation operators 
$\hat{z}$ which explains the notion ``coherent state''.
As is well known, property (\ref{2.6}) implies that the uncertainty 
relation for the self adjoint operators 
\be \label{2.7}
\hat{x}:=[\hat{z}+\hat{z}^\dagger]/2,\;\;
\hat{y}:=-i[\hat{z}-\hat{z}^\dagger]/2
\ee
is saturated, that is
\be \label{2.8}
[<\hat{x}^2>_z-(<\hat{x}>_z)^2]\;
[<\hat{y}^2>_z-(<\hat{y}>_z)^2]=\frac{1}{4}\;|<[\hat{x},\hat{y}]>_z|
\ee
where $<.>_z:=<\psi_z,.\;\psi_z>/||\psi_z||^2$ denotes the expectation 
value with respect to $\psi_z$ (notice that $\psi_z$ is in general not 
automatically normalized). This is a second property commonly attributed 
to coherent states \cite{31}. 

Finally, under certain technical assumptions spelled out in \cite{32}
the completeness relation
\be \label{2.9}
1_{{\cal H}}=\int_{\overline{{\cal C}}} \; d\mu(q_0)\; \delta_{q_0}\;  
\delta_{q_0}[.]
\ee
implies that there exists a measure $\nu$ on $\overline{{\cal 
C}}^{\mathbb{C}}$ such that
\be \label{2.10}
1_{{\cal H}}=\int_{\overline{{\cal C}}^{\mathbb{C}}} \; d\nu(z)\; 
\psi_z\; <\psi_z,.>
\ee
This concludes the general discussion. The interested reader may verify
\cite{21,23} 
that the coherent states for Maxwell Theory on Minkowski space result 
from the complexifier
\be \label{2.11}
C=\frac{1}{2\kappa^2}\int_{\mathbb{R}^3}\;d^3x\; \delta_{ab} E^a 
\sqrt{-\Delta}^{-1} E^b
\ee
where $E^a$ is the Maxwell electric field, $\Delta$ is the 
Laplacian on $\mathbb{R}^3$, $\kappa$ is the electric 
charge and 
$\alpha=\hbar\kappa^2$ is the Feinstruktur constant.

\subsection{Complexifiers for Background Independent Gauge Theories}
\label{s2.2}

As explained in detail in \cite{33,34}, gauge theories with 
compact gauge group $G$ provide 
an almost perfect arena for the general theory summarized in the 
previous 
subsection. Let us explain this in some detail:
\begin{itemize}
\item[1.] {\it Classical Phase Space}\\
The role of $\cal C$ is 
played by some space of smooth 
connections $\cal A$ over some $D-$dimensional spatial manifold 
$\sigma$. The role of 
$\cal M$ is then simply $T^\ast {\cal A}$. The configuration and 
momentum coordinates on this phase space then are simply real 
valued connection one 
forms $A_a^j(x)$ (potentials) and Lie algebra valued vector densities
$E^a_j(x)$ (electric fields) respectively which enjoy the following 
Poisson brackets
\be \label{2.12}
\{A_a^j(x),A_b^k(y)\}=
\{E^a_j(x),E^b_k(y)\}=0,\;\;
\{E^a_j(x),A_b^k(y)\}=\kappa\;\delta^a_b\;\delta_j^k\;
\delta(x,y)
\ee
Here $\kappa$ denotes the square of the coupling constant of the gauge 
theory, $a,b,c,..=1,..,D$ denote spatial tensor indices and 
$j,k,l,..=1,..,\dim(G)$ denote Lie algebra indices. We will assume 
that $G$ is connected, semisimple and take the convention that the 
internal metric is 
just $\delta_{jk}$. 
\item[2.] {\it Distributional Configuration Space}\\
Now consider
arbitrary, finite piecewise analytic (more precisely semianalytic 
\cite{14}) graphs embedded in $\sigma$ which we think of as collections 
of edges $e$, that is, piecewise analytic one dimensional paths which 
intersect at most in their endpoints, called the set $V(\gamma)$ of 
verticies of $\gamma$. 
Denote by $E(\gamma)$ the set of edges of $\gamma$. Given $\gamma$
consider functions {\it cylindrical over $\gamma$} of the form
\be \label{2.11a}
f:\;{\cal A}\to \mathbb{C};\;
A\mapsto f(A)=f_\gamma(\{A(e)\}_{e\in E(\gamma)})
\ee
where $f_\gamma:\;G^{|E(\gamma)|} \to \mathbb{C}$ is a complex valued 
function on $|E(\gamma|$ copies of $G$ and $A(e)$ denotes the holonomy
of $A$ along $e$. Functions of the form (\ref{2.11a}) form an Abelian 
$^\ast$ algebra under pointwise operations with the involution given by 
complex conjugation. We can turn it into an Abelian 
$C^\ast-$algebra,
usually called Cyl (cylinder functions) with respect to the sup norm on 
$\cal A$ that is
\be \label{2.12a}
||f||:=\sup_{A\in {\cal A}} \;|f(A)|
\ee
As is well known \cite{35}, Abelian $C^\ast-$algebras $\mathfrak{A}$ are 
isometric isomorphic to another Abelian $C^\ast-$algebra which consists 
of continuous functions on a compact Hausdorff space 
$\Delta(\mathfrak{A})$, called the spectrum of $\mathfrak{A}$. Denote 
the spectrum of Cyl by $\overline{{\cal A}}$. It has a nice geometrical 
interpretation as a space of generalized connections in the sense that
the holonomy of 
$A\in \overline{{\cal A}}$ satisfies all the usual algebraic relations 
satisfied by smooth holonomies, that is 
$A(e\circ e')=A(e) A(e')$ if the end point of $e$ is the beginning point 
of $e'$ and $A(e^{-1})=(A(e))^{-1}$ ,
but that smoothness or even continuity is no longer required. The 
topology on $\overline{{\cal A}}$ is the Gel'fand topology which in 
this case boils down to saying that a net of generalized connections 
converges when the corresponding net of holonomies for all 
possible paths converges. See 
\cite{13,36,1b} for more details.
\item[3.] {\it Hilbert Space}\\
Being a compact Hausdorff space, a natural set of representations
of the Poisson$^\ast-$algebra generated by all the holonomies and all 
the electric fluxes through co-dimension 1 (piecewise analytic) surfaces 
$S$ 
\be \label{2.13}
E_j(S):=\int_S \; \ast E_j
\ee
where $\ast E_j$ denotes the pseudo $(D-1)$form dual to $E^a_j$,     
should be of the form ${\cal H}=L_2(\overline{{\cal A}},d\mu)$ where 
$\mu$ is a Borel probability measure. It turns out that all cyclic
representations that carry a unitary representation of the 
diffeomorphism group Diff$(\sigma)$ are of this form \cite{14} and the 
corresponding measure is unique and was first discovered in \cite{13}.
See e.g. \cite{1b} for all details. For our purposes it is enough to  
know that $\cal H$ admits a natural orthonormal basis, called spin 
network functions (SNWF). They are labelled by a graph $\gamma$, a 
collection $\pi=\{\pi_e\}_{e\in E(\gamma)}$ of irreducible, non trivial
representations 
of $G$ and collections $m=\{m_e\}_{e\in E(\gamma)},\;\;n=\{n_e\}_{e\in 
E(\gamma)}$ of integers $m_e,\; n_e=1,.., \dim(\pi_e)$  labelling
the edges of $\gamma$ and are explicitly defined by
\be \label{2.14}
T_{\gamma,\pi,m,n}(A):=\prod_{e\in E(\gamma)}\; \sqrt{\dim(\pi_e)}\;
[\pi_e(A(e))]_{m_e n_e}
\ee
\end{itemize}
In order to further apply the general theory of the previous subsection 
we must provide a complexifier. The complexifier for Maxwell theory
displayed in (\ref{2.11}) is motivated by the fact that the associated 
annihilation operators are precisely those that enter the Maxwell 
Hamiltonian. In General Relativity there is no a priori Hamiltonian
but there is the Hamiltonian constraint. Hence one might be tempted 
to choose a complexifier whose associated annihilation operator is 
related to the Hamiltonian constraint. Unfortunately, the Hamiltonian 
constraint is, in contrast to Maxwell theory, neither polynomial nor 
does it have a quadratic piece with respect to which a perturbation 
scheme can be defined. Hence, the notion of an annihilation 
operator defined by the Hamiltonian constraint is ill 
defined\footnote{The situation slightly improves when a physical 
Hamiltonian is available, see \cite{5}.}. On the other hand, since we 
here just want to construct coherent states which approximate well our
elementary holonomy and flux operators which are defined on the 
kinematical Hilbert space (on which the Hamiltonian constraint is not 
satisfied) in order to verify whether other kinematical operators (i.e. 
not 
invariant under the gauge motions generated by the spatial 
diffeomorphism and Hamiltonian constraints) such as the volume operator    
have been correctly quantized, the motivation to use the Hamiltonian 
constraint as a selection criterion for the complexifier is anyway 
less motivated.

In lack of a better selection criterion, we take here a practical 
attitude and would like to consider a complexifier which comes close 
to the Maxwell one (\ref{2.11}) which obviously satisfies all 
the requirements of definition \ref{def2.1}. Since we have 
applications in General Relativity in mind, we must preserve background 
independence and therefore the Minkowski background Laplacian entering
(\ref{2.11}) must be replaced by something background metric independent. 
One possibility is to use a background independent Laplacian which 
depends on the dynamical 3-metric of $q_{ab}$ which 
$E^a_j/\sqrt{|\det(E)|}$
is its triad. However, this would lead to a very complicated object
with which no practical calculations are possible. In fact, the 
practical use of coherent states in Maxwell theory rests on the fact 
that (\ref{2.11}) is quadratic in the momenta (electric fields) which 
leads to states which are basically Gaussians in both the position and 
the momentum representation.
This motivates to keep our complexifier quadratic in the momenta as 
well. Furthermore, we must preserve $G$ invariance. For Abelian 
gauge theories the electric fields are already gauge invariant but 
not for non Abelian gauge theories.

Thus a first attempt would be to define as complexifier
\be \label{2.15}
C\propto \int_\sigma \; d^3x\; q_{ab} \; E^a_j E^b_k 
\delta^{jk}/\sqrt{\det(q)}
\ee
where had to replace the background metric $\delta_{ab}$ in (\ref{2.11})
by the dynamical metric and in order to make (\ref{2.15}) spatially 
diffeomorphism invariant we have included a density factor 
$1/\sqrt{\det(q)}$. However, it is easy to see that (\ref{2.15}) becomes 
\be \label{2.16} 
C\propto V=\int_\sigma \; d^3x\; \sqrt{|\det(E)|}
\ee
the volume functional. While it satisfies the requirements of a 
complexifier, and admits a quantization as a positive self adjoint 
operator, its spectral decomposition is not analytically available
so that $C=V$ is not practically useful.

Hence, what we need is a gauge invariant, background independent 
expression, quadratic in the
electric fields 
which preferably is non vanishing everywhere on $\sigma$ and 
which can be expressed in terms of (limits of) electric 
fluxes since 
only those are well defined in the quantum theory. In \cite{21} it is 
shown that in non Abelian gauge theories no quadratic 
complexifier based strictly on fluxes exists that meets all these 
requirements. The way out is to give up the the requirement that the 
complexifier is composed out of the fluxes but to allow more general 
objects than fluxes. There are basically two proposals in the literature.
The first \cite{10} replaces fluxes by gauge covariant fluxes. The second 
\cite{21} replaces the flux by areas. We will review these two proposals 
separately.

In what follows we assume that as in General Relativity the canonical 
dimension of $E^a_j$ is cm$^0$ and that of $A_a^j$ is cm$^{-1}$ so that 
(\ref{2.17}) has dimension 
cm$^{D-1}$. Since the kinetic term in the canonical action is
$\int_{\mathbb{R}} dt \int_\sigma d^D E^a_j \dot{A}_a^j/\kappa$ 
it follows that $\hbar\kappa$ has dimension cm$^{D-1}$.

\subsubsection{Gauge Covariant Flux Complexifiers}
\label{s2.2.1}

Given a surface $S$ select a point $p(S)\in S$.
Furthermore, for each point 
$x\in S$ choose a path $\rho_S(x)\subset S$ within $S$ with beginning 
point $p(S)$ and ending point $x$. Denote the path system by ${\cal 
P}_S$. 
The gauge covariant flux of $E$ through 
$S$ subordinate to the path system ${\cal P}_S$ and the edge $e_S$ is 
defined by 
\be \label{2.16a}
E_j(S)\tau_j:=\int_S\; 
{\rm Ad}_{A(\rho_S(x))}((\ast E)(x))
\ee
Here $i\tau_j$ are the Pauli matrices, $\ast E=\frac{1}{2}\epsilon_{abc}
dx^a\wedge dx^b \; E^c_j \tau_j$ and Ad denotes the adjoint action of $G$ 
on its Lie algebra. Obviously, (\ref{2.16a}) transforms in the adjoint 
representation under gauge transformations at $p(S)$.

Let $\cal S$ be a collection of surfaces with associated path systems 
${\cal P}_S$ for each $S\in{\cal S}$ and 
$\mu$ a measure on $\cal S$. Let $K$ be any positive definite, measurable  
function on 
$\cal S$.
A gauge covariant flux complexifier (GCFC) is defined by 
\be \label{2.16b}
C:=\frac{1}{2 L^{D-1} \kappa} \int_{{\cal S}}\; d\mu(S) \;K(S)\;
[-\frac{1}{2}{\rm Tr}(E(S)^2)]
\ee
which is manifestly gauge invariant (one could absorb $K,\;L$ into $\mu$).
Here $L$ is a parameter of dimension of length and we assume both 
$\mu,\;K$ to be dimensionless.

The mostly studied case is when $D=3$ and $\cal S=C_2(P)$ is a discrete 
set of 
oriented surfaces 
which coincide with the faces (its sub 2-complex) of a polyhedronal cell 
partition $P$ of 
$\sigma$. In this case $\mu$ is just the counting measure and for 
convenience one chooses $K=1$. 
We will denote the associated complexifier by $C_P$. In this case the 
complexified connection is given by
\be \label{2.16c}
Z_a^j(x)=A_a^j(x)-\frac{i}{L^2}\sum_{S\in C_2(P)} 
\int_S\; \frac{1}{2}\epsilon_{abc} dy^b\wedge dy^c \;\
[{\rm Tr}(E(S)\;{\rm Ad}_{\rho_S(x)}(\tau_j))]
\; \delta(x,y)
\ee
Notice that the series involved in $Z_a^j$ terminates at the first term.
This is because when computing the second iterated Poisson bracket there
is a double sum over surfaces involved but 
because the paths 
$\rho_S(x)$ are disjoint from $S'$ for $S'\not= S$ 
there is no contribution from $S'\not=S$ to 
$\{C_P,A_a^j(x)\}_{(2)}$. For $S'=S$ there is in principle a contribution 
but by the regularization \cite{1b} the classical flux does not 
Poisson act on paths lying in its associated surface.

This connection is distributional but fortunately we are only 
interested 
in the integral of (\ref{2.16c}) over one dimensional paths $e$ given by
\be \label{2.16e}
i L^2\int_e dx^a[Z_a^j(x)-A_a^j(x)]=\sum_{S\in C_2(P)}\;\sum_{x\in S\cap 
e} \; \sigma_x(S,e)
[{\rm Tr}(E(S)\;{\rm Ad}_{\rho_S(x)}(\tau_j))]
\ee
where 
\be \label{2.16f}
\sigma_z(S,e)=\frac{1}{2} \int_e\; dx^a\; \epsilon_{abc}\; \int_S \; 
dy^b\wedge dy^c \delta(x,y) \delta_{x,z}
\ee
is the signed intersection number at $z\in e\cap S$ which here we have 
assumed to be an interior point (otherwise there is an additional factor 
of $1/2$, in \cite{1b}).

\subsubsection{Area Compexifier}
\label{s2.2.2}

Let $\cal S$ be a collection of surfaces, $\mu$ a measure on $\cal S$ and 
$K(S,S')$ a positive definite integral kernel. An area complexifier 
is given by the expression 
\be \label{2.16g}
C=\frac{1}{a^{D-1} \kappa} 
\int_{{\cal S}}\; d\mu(S) \;
\int_{{\cal S}}\; d\mu(S) \; K(S,S')\; {\rm Ar}(S)\; {\rm Ar}(S')
\ee
where $a$ is a parameter of dimension of length. Here Ar$(S)$ is 
the gauge invariant ``modulus of the electric flux''
\be \label{2.17}
{\rm Ar}(S):=\int_S \; \sqrt{{\rm Tr}([\ast E]^2)}
\ee
which in General Relativity has the meaning of the {\it area of S}.

The most studied case arises from a diagonal and constant integral kernel
and the following choices of $\cal S$ and $\mu$ respectively.
\begin{Definition} \label{def2.2} ~\\
i.\\
A stack $s$ in $\sigma$ is a D-dimensional submanifold with the topology
of $\mathbb{R}\times (0,1]^{D-1}$.\\
ii.\\
A stack family $S=\{s_\alpha\}$ is a partition of $\sigma$ into stacks 
which are mutually disjoint.\\
iii.\\
D families of foliations $F_I,\;I=1,..,D$ of $\sigma$ generated by 
vector fields
$\partial/\partial t^I,\;I=1,..,D$ are said to be linearly independent 
if the vector fields $\partial/\partial t^I$ are everywhere linearly
independent.\\
iv.\\
D stack families $S^I$ are said to be linearly independent provided that 
there exist D linearly independent foliations $F_I$ 
such that the leaves of the foliation 
$F_I$ is transverse to every 
stack in $S^I$. That is, the intersection $s^I_{\alpha t}$ of any leaf 
$L_{It},\;t\in 
\mathbb{R}$ of $F_I$ 
with any stack $s^I_\alpha$ in $S^I$, called a plaquette, has topology 
$(0,1]^{D-1}$.\\ 
v.
The collection of the plaquettes $s^I{\alpha t}$ is called a 
parquette at time t within $L_{It}$.
\end{Definition}
In general $\sigma$ will have to be partitioned into pieces that admit 
$D$ linearly independent foliations each. We construct the 
complexifier for one such piece below, the complete complexifier is 
then the sum over the pieces. 

The complexifier defined by $D$ linearly independent stack families 
is now is defined by
\be \label{2.18}
C:=\frac{1}{2\kappa a^{D-1}} \sum_{I=1}^D \sum_\alpha 
\int_{\mathbb{R}}\; 
dt\;
[{\rm Ar}(p^I_{\alpha t})]^2  
\ee
Here we take the foliation parameter $t$ to be dimensionless, $a$ is 
a parameter with dimension cm$^1$ so that $C/\hbar$ is dimensionfree
and $p^I_{\alpha t}=s^I_\alpha \cap L_{It}$ denotes the plaquette at 
time $t$ within the stack $s^I_\alpha$ in direction $I$.
For Abelian gauge theories also the following simpler expression is 
available      
\be \label{2.18a}
C:=\frac{1}{2\kappa a^{D-1}} \sum_{I=1}^D \sum_\alpha 
\int_{\mathbb{R}}\; 
dt\; [E_j(p^I_{\alpha t})]^2  
\ee
which uses the gauge invariant flux rather than the areas.

Let us now compute the complexified connections. Notice that due to the 
fact that each stack is foliated by squares with half open and half 
closed boundaries, for each $x\in \sigma$ and each direction $I$ there 
exists a unique stack $s^I_\alpha(x)$ corresponding to a label 
$\alpha_I(x)$ such that $x\in s^I_\alpha$. 
Likewise, for each direction $I$ there exists a unique leaf $L_It(x)$ 
corresponding to a time $t_I(x)$ such that $x\in L_{It}$. Consider 
the one parameter family of embeddings $X^I_{\alpha t}:\;[0,1)^{D-1}\to 
p^I_{\alpha t}$, then there exists a unique $u_I(x)$ such that 
$x=X^I_{\alpha_I(x) t_I(x)}(u_I(x))$. We now set
\ba \label{2.19}
J_I(x) &:=& |\det(\frac{\partial X^I_{\alpha t}(u)}{\partial (t, 
u)})|_{\alpha=\alpha_I(x),t=t_I(x),u=u_I(x)}
\nonumber\\
n_a^I(x) &:=& \frac{1}{(D-1)!} \epsilon_{ab_1 .. b_{D-1}}\; 
\epsilon_{l_1 
.. 
l_{D-1}} 
\frac{\partial X^{I b_1}_{\alpha t}(u)}{\partial u^{l_1}} ..
\frac{\partial X^{I b_{D-1}}_{\alpha t}(u)}{\partial u^{l_{D-1}}}
\ea
For the non Abelian
complexifier we find 
\be \label{2.20}
Z_a^j(x)=A_a^j(x)-\frac{i}{a^{D-1}} E^b_j(x)\sum_I \frac{n_b^I(x) 
n_a^I(x)}{J_I(x)}\;
\frac{{\rm Ar}(p^I_{\alpha_I(x) t_I(x)})}{\sqrt{[E^c_k(x) n_c^I(x)]^2}} 
\ee
while for the Abelian one we obtain 
\be \label{2.21}
Z_a^j(x)=A_a^j(x)-\frac{i}{a^{D-1}} \sum_I \frac{n_a^I(x)}{J_I(x)}\;
E_j(p^I_{\alpha_I(x) t_I(x)})
\ee
Notice that in both cases the imaginary part of $Z_a^j$ is only quasi 
local in $E^a_j$, that is, we can recover $E^a_j$ from $Z_a^j$ only up 
to the resolution provided by the parquettes.

\subsection{Coherent States for Background Independent Gauge Theories}
\label{s2.3}

We now come to compute the coherent states. The first step is to write 
the $\delta$ distribution as 
\be \label{2.22}
\delta_{A_0}=\sum_s \; T_s(A_0)\;  <T_s,.>
\ee
where the sum is over all spin network labels $s=(\gamma,\pi,m,n)$. 
Hence the coherent state is given by
\be \label{2.23}
\psi_Z=\sum_s \; T_s(Z) \; <e^{-\hat{C}/\hbar} T_s,.>
\ee
Here $\hat{C}$ is obtained by replacing in (\ref{2.16a}), (\ref{2.18}) 
or 
(\ref{2.18a}) respectively the gauge covariant flux, area or flux 
functionals by the gauge covariant flux, area or flux operator 
\cite{10,7,37} respectively which are 
positive, self adjoint operators with pure point spectrum only. 

It remains to compute the action of the heat kernel and for this purpose 
we restrict to the case $D=3$ of ultimate interest. Again we do this 
separately for the two types of complexifiers.

\subsubsection{Gauge Covariant Flux Coherent States}
\label{s2.3.1}

There is in principle an operator ordering problem involved in the 
quantization of (\ref{2.16a}), however, the regularization of \cite{1b} 
shows that there is no action of the operator valued distribution 
$\ast E(x)$ on a holonomy $A(p)$ if $\ast E(x)$ is smeared over an 
infinitesimal surface element of a surface in which the path $p$ lies.
Let us introduce the matrices 
\be \label{2.23a}
O_{jk}(g):=-\frac{1}{2}\;{\rm Tr}(\tau_k {\rm Ad}_g(\tau_j))
\ee
where we have assumed the normalization ${\rm Tr}(\tau_j 
\tau_k)=-2\delta_{jk})$. Since $G$ is compact, we can always embed into 
a subgroup of some $U(N)$ so that 
$\overline{\tau_j^T}=-\tau_j,\;\overline{g}^T=g^{-1}$ whence $O_{jk}(g)$ 
is real valued. Moreover, the obvious identity 
$O_{jk}(g)=O_{kj}(g^{-1})$ as well as the fact that Ad acts on Lie$(G)$
whence $Ad_g(\tau_j)=O_{jk}(g)\tau _k$ reveals that 
\be \label{2.23b}
O_{jk}(g) O_{jl}(g)=\delta_{kl}  
\ee
so that $g\mapsto O_{jk}(g)$ is a subgroup of $O(\dim(G))$. 

The known quantization of the non gauge covariant flux \cite{1b,37} 
together with the above mentioned trivial action on 
$O_{jk}(A(\rho_S(x))$ now reveal that 
\be \label{2.23c}
\widehat{E_j(S)} T_{\gamma,j,m,n} = i\ell_P^2 \sum_{e\in E(\gamma)} \;
\sum_{x\in S\cap e}\; \sigma_x(S,e) \; O_{jk}(A(\rho_S(x)))\; 
\frac{1}{4} X^k_e \; T_{\gamma,j,m,n}
\ee
where $X^k_e$ is the right invariant vector field of $G$ acting on 
$g=A(e)$, specifically $X^k_e={\rm Tr}(\tau_j g \partial/\partial g^T)$.
Here we have assumed that the graph has been adapted to $S$ by suitable 
subdivisions of edges, such that each edge of $\gamma$ is either 
outgoing from an 
isolated 
intersection point or completely lies within $S$ or lies completely 
outside $S$. 

Formula (\ref{2.23c}) can now be plugged into (\ref{2.16b}). Since again 
there is no action of $widehat{E(S)}$ on $\rho_S(x)$ we find 
\be \label{2.23d}
\widehat{E_j(S)}^2 T_{\gamma,j,m,n} = -\ell_P^4 \sum_{e,e'\in E(\gamma)} 
\;
\sum_{x\in S\cap e}\; \sigma_x(S,e) \;
\sum_{y\in S\cap e'}\; \sigma_y(S,e') 
\; O_{kl}(A(\rho_S(x)^{-1}\circ \rho_S(y)))\; 
\frac{1}{16} X^k_e X^l_{e'} \; T_{\gamma,j,m,n}
\ee

The appearance of the matrix $O_{kl}(A(\rho_S(x)^{-1}\circ \rho_S(y)))$
makes the computation of the spectrum of (\ref{2.23d}) rather difficult 
for a general graph. However, it becomes simple in case that the 
graph is such that the surface $S$ has only a single isolated 
intersection point $x$ with the graph. In that case (\ref{2.23d}) 
becomes
\be \label{2.23e}
\widehat{E_j(S)}^2 T_{\gamma,j,m,n} = -\ell_P^4 
[\sum_{e\in E(\gamma)} \;\sum_{x\in S\cap e}\; \sigma_x(S,e) \;
\frac{1}{4} X^j_e]^2 \; T_{\gamma,j,m,n}
\ee
One can now introduce similar as in \cite{37} the vector fields 
\be \label{2.23f}
Y_S^{j \pm}=-i\sum_{\sigma_x(e,S)=\pm 1} X^j_e/2,\;   
Y_S^j=Y^{j +}_S+Y^{j -}_S
\ee
so that we obtain the linear combinations of Casimir operators   
\be \label{2.23g}
\widehat{E_j(S)}^2 T_{\gamma,j,m,n} = \frac{\ell_P^4}{4}
[2 (Y^{j +}_S)^2+2 (Y^{j -}_S)^2- (Y^j_S)^2] \; T_{\gamma,j,m,n}
\ee
A special case arises when $x$ is an interior point of a single edge
$e=(e_1)^{-1}\circ e_2$ intersected transversely so that 
$\sigma_x(S,e_1)=-\sigma_x(S,e_2)=\pm 1$ and thus $T_{\gamma, j, 
m,n}$ is gauge invariant at $x$. Then (\ref{2.23g} further simplifies to
\be \label{2.23h}
\widehat{E_j(S)}^2 T_{\gamma,j,m,n} = \ell_P^4\;
[-i X^j_e/2]^2 \; T_{\gamma,j,m,n}
\ee
For $G=U(1)^3$ or $G=SU(2)$ respectively the eigenvalues of 
$(-iX^j_e)^2$ are given by $(n^j_e)^2$ and $j_e(j_e+1)$ respectively.
This special case arises for the case of the polyhedronal cell complex 
complexifier when $\gamma$ is a graph dual to it, that is, there is 
precisely one edge $e$ of $\gamma$ which intersects a given face $S$ 
and if so transversely. 

\subsubsection{Area Coherent States}
\label{s2.3.2}

Notice
that for each direction $I$, each graph $\gamma$ and each stack $\alpha$ 
the Lebesgue measure 
of the set of times $t$ such that $p^I_{\alpha t}$ contains a vertex of 
$\gamma$ or that $p^I_{\alpha t}$ contains entire segments of edges of 
$\gamma$ vanishes. By the properties of the area operator and flux 
operator, it follows that those points do not contribute to the heat 
kernel evolution and therefore we may assume without loss of generality 
that each $p^I_{\alpha t}$ intersects the edges of $\gamma$ at most 
transversely in an interior point. Now consider in the non Abelian 
case for natural numbers 
$N_e\in \mathbb{N}_0$ the set 
\be \label{2.24}
S^{I \alpha \gamma}_N:=\{t\in \mathbb{R};\;
|p^I_{\alpha t}\cap e|=N_e\}
\ee
where we have abbreviated $N:=\{N_e\}_{e\in E(\gamma)}$. This is the set 
of parquettes within stack $s^I_\alpha$ which intersect edge $e$ 
precisely $N_e$ times transversely. 
Likewise, consider for the Abelian case for integers $N_e \in 
\mathbb{Z}$ 
\be \label{2.25}
S^{I \alpha \gamma}_N:=\{t\in \mathbb{R};\;
\sum_{x\in p^I_{\alpha t}\cap e} \sigma(p^I_{\alpha,t},e)_p=N_e\}
\ee
where for any surface $S$ intersecting $e$ transversely, the number 
$\sigma(S,e)_p$ for $p\in S\cap e$ takes the value $+1$ or $-1$ 
respectively if the 
orientations of $S$ and $e$ at $p$ agree or disagree respectively.
This is the set of parquettes within stack $p^I_\alpha$ whose signed 
intersection number with edge $e$ is precisely $N_e$.

In both cases let
\be \label{2.26}
l^{I\alpha \gamma}_N:=\int_{S^{I \alpha}_N}\; dt
\ee
be the Lebesgue measure or {\it length} of those sets. These length 
functions are needed in order to define a cylindrically consistent 
family of heat kernels as was first observed in \cite{34}. Then 
the action of the complexifier on SNWF's is diagonal
\be \label{2.27}
\frac{\hat{C}}{\hbar} T_s =\lambda_s T_s
\ee
The corresponding eigenvalues are given for $G=SU(2)$ by
\be \label{2.28}
\lambda_s=\frac{\ell_P^2}{2a^2}\sum_{I,\alpha}\; \sum_N\; 
l^{I \alpha \gamma}_N 
\; [\sum_{e\in E(\gamma)}\; N_e\; \sqrt{j_e(j_e+1)}]^2
\ee
while for $G=U(1)^3$ they are given by 
\be \label{2.29}
\lambda_s=\frac{\ell_P^2}{2a^2}\sum_{I,\alpha}\; \sum_N\; 
l^{I \alpha \gamma}_N 
\; [\sum_{e\in E(\gamma),j}\; N_e\; n^j_e]^2
\ee
Here we have used that the irreducible, non trivial representations of 
$SU(2)$ 
are given by positive, half integral spin quantum numbers $j_e\not=0$ 
while for 
$U(1)^3$
they are given by triples of integers $n_e^j\not=0,\;j=1,2,3$. 
Furthermore, with $\kappa=8\pi G_{{\rm Newton}}$, $\ell_P^2=\hbar 
\kappa$ is the Planck area. The ratio $t:=\ell_P^2/a^2$ is known as the 
classicality parameter. Without dynamical input, this is a free 
parameter for our coherent states that decides up to which scale 
the fluctuations of operators are negligible.

\subsection{Cut -- Off Coherent States}
\label{s2.4}

Formulae (\ref{2.23}), (\ref{2.23d}) (\ref{2.28}) and (\ref{2.29}) 
display the 
coherent states in closed form. Unfortunately, although the eigenvalues 
of the heat kernel grow quadratically with the representation weight,
these states are still not normalizable because the Hilbert space is not 
separable, or in other words, the SNWF's are labelled by the continuous 
parameter $\gamma$. In view of the uniqueness result when 
insisting on background independence, the non separability is 
not avoidable and one must accept it. The observation is 
that (\ref{2.23}) defines a 
well defined 
distribution on the dense subset of $\cal H$ consisting of the finite 
linear span of SNWF's. To 
extract normalizable information from $\psi_Z$ we introduce the notion 
of a cut -- off state labelled by a graph $\gamma$. These are defined by
\be \label{2.30}
\psi_{Z,\gamma}:=\sum_{s; \gamma(s)\subset \gamma}\; T_s(Z)\; 
<e^{-\hat{C}/\hbar} T_s,.>
\ee
That is, the sum over all spin networks $s=(\gamma(s),\pi(s),m(s),n(s))$ 
is truncated or {\it cut off} to those whose
graph entry $\gamma(s)$ is a subgraph of the given $\gamma$. The Ansatz 
is then to use $\psi_{Z,\gamma}$ for suitable $\gamma$ as a 
semiclassical 
state.    

Notice that both (\ref{2.28}) and (\ref{2.29}) respectively can be 
rewritten in the form
\be \label{2.31}
\lambda_s=\frac{t}{2}\sum_{e,e'}\; l^\gamma_{e,e'}\; 
\sqrt{j_e(j_e+1)}\; \sqrt{j_{e'}(j_{e'}+1)}
\ee
and 
\be \label{2.32}
\lambda_s=\frac{t}{2}\sum_{e,e'}\; l^\gamma_{e,e'}\; 
n^j_e\; n^j_{e'}
\ee
where the {\it edge metric} 
\be \label{2.32a}
l^\gamma_{e,e'}=
\sum_{I,\alpha}\; \sum_N\; 
l^{I \alpha \gamma}_N\; N_e N_{e'}
\ee
has entered the stage. Such non diagonal edge metrics have already 
appeared in other background dependent contexts \cite{23,24}. 
The edge metric decays quickly off 
the diagonal because for most edge pairs $e\not=e'$ there is no 
direction 
and no stack in that direction intersecting both $e,e'$ which means 
that $l^{I\alpha \gamma}_N=0$ for $N_e,\;N_{e'}\not=0$ for such edge 
pairs. It is for this 
reason that we will be able to actually carry out our calculations.   

Using the edge metric, formulas (\ref{2.20}), (\ref{2.21}) and 
(\ref{2.28}), (\ref{2.29}) admit an interesting reformulation: \\
The {\it signed intersection number} between a path $e$ and a surfaces 
$S$ is defined by (adopting convenient parametrization)
\ba \label{2.32b}
\sigma(S,e) &:=& \int_e dx^a\;\int_S \; dy^b \; dy^c\; 
\frac{1}{2}\;\epsilon_{abc}
 \;
\delta(x,y)
=\int_0^1\; dt\int_{[0,1]^2}\; d^2u\;
[\epsilon_{abc} \dot{e}^a(t)
\frac{\partial S^b(u)}{\partial u^1}
\frac{\partial S^c(u)}{\partial u^2}]\; \delta(e(t),S(u))
\nonumber\\
&=& \sum_{x\in S\cap e}\; \sigma_x(S,e)
\ea
while the {\it intersection number} is given by
\be \label{2.32c}   
|\sigma|(S,e):=
=\int_0^1\; dt\int_{[0,1]^2}\; d^2u\;
|\epsilon_{abc} \dot{e}^a(t)
\frac{\partial S^b(u)}{\partial u^1}
\frac{\partial S^c(u)}{\partial u^2}|\; \delta(e(t),S(u))
\ee
Both expressions can be regularized in such a way that entire segments 
of $e$ that lie inside $S$ do not contribute to the integral \cite{37}.
Notice that $|\sigma|(e,S)\not=|\sigma(e,S)|$. Then it is not difficult 
to see that for $SU(2)$
\be \label{2.32d}
l^\gamma_{e,e'}=\sum_{\alpha,I} \; \int \; dt\; 
|\sigma|(e,p^{\alpha I}_t)\; |\sigma|(e',p^{\alpha I}_t)
\ee
while for $U(1)^3$ 
\be \label{2.32e}
l^\gamma_{e,e'}=\sum_{\alpha,I} \; \int \; dt\; 
\sigma(e,p^{\alpha I}_t)\; \sigma(e',p^{\alpha I}_t)
\ee
To verify (\ref{2.32d}), (\ref{2.32e}) it is easiest to use directly
the action of non Abelian area and Abelian flux operators respectively 
on the corresponding SNWF \cite{37} (with only transverse 
intersections)
\ba \label{2.32f}
{\rm Ar}(S) T_{\gamma,j,m,n} &=& \ell_P^2 \;[\sum_{e\in E(\gamma)}\; 
|\sigma|(e,S)\; \sqrt{j_e(j_e+1)}]\; T_{\gamma,j,m,n}
\nonumber\\
E_j(S) T_{\gamma,n} &=& \ell_P^2 \;[\sum_{e\in E(\gamma)}\; 
\sigma(e,S)\; n_e^j]\; T_{\gamma,n}
\ea 
and to plug this formula into the expression for $C$. An alternative 
proof is by realizing that in the non Abelian or Abelian case 
respectively 
\be \label{2.32g}
\chi_{S^{\alpha I}_{N}}(t)=\prod_{e\in E(\gamma)} 
\delta_{|\sigma|(p^{\alpha I}_t,e),N_e},\;\;
\chi_{S^{\alpha I}_{N}}(t)=\prod_{e\in E(\gamma)} 
\delta_{\sigma(p^{\alpha I}_t,e),N_e}
\ee
where $\chi_S$ denotes the characteristic function of a set. When 
plugging (\ref{2.32g}) into (\ref{2.32a}) and solving the Kronecker 
$\delta$'s when carrying out the sum over the integers $N$, 
(\ref{2.32d}) and (\ref{2.32e}) respectively result. 

From the easily verifiable properties of the (signed) intersection 
numbers
\be \label{2.32h}
\sigma(e\circ e',S)=\sigma(e,S)+\sigma(e',S),\;
\sigma(e^{-1},S)=-\sigma(e,S);\;\;
|\sigma|(e\circ e',S)=|\sigma|(e,S)+|\sigma|(e',S),\;
|\sigma|(e^{-1},S)=|\sigma|(e,S)
\ee
it follows immediately that 
\be \label{2.32i}
l^\gamma(e\circ e',e\circ 
e')=l^\gamma(e,e)+l^\gamma(e',e')+2l^\gamma(e,e'),\;\;
l^\gamma(e^{-1},e^{-1})=l^\gamma(e,e)
\ee
This is precisely the generalization to non diagonal edge metrics
of the cylindrical consistency conditions of the complexifier 
\cite{21,34}. Notice that for the general area complexifier 
(\ref{2.16g}) 
we arrive instead at the edge metrics
\be \label{2.32l}
l^\gamma_{e,e'} =\int_{{\cal S}}\;d\mu(S)\; \int_{{\cal S}}\;d\mu(S') \;
|\sigma|(S,e)\;K(S,S')\; |\sigma|(S',e'),\;\;\;
l^\gamma_{e,e'} =\int_{{\cal S}}\;d\mu(S) \;\int_{{\cal S}}\;d\mu(S) \; 
\sigma(S,e)\;K(S,S')\; \sigma(S',e')
\ee

Finally we have for any edge $e$
\be \label{2.32j}
\int_e\;  dx^a\; ia^2[Z_a^j-A_a^j](x) = \sum_{I,\alpha} \int\; dt\; 
{\rm Ar}(p^{I \alpha}_t)
\int_0^1 \; ds\;
\frac{(n_c^I E^c_j)(e(s))}{\sqrt{[(n_b^I E^b_j)(e(s))]^2}}\;
\; \int d^2u \;[\dot{e}^a(s) n_a^{\alpha I t}(u) \delta(p^{\alpha 
I}_t(u),e(s))]
\ee
in the non Abelian case while for the Abelian case 
\be \label{2.32k}
\int_e\;  dx^a\; ia^2[Z_a^j-A_a^j](x) = \sum_{I,\alpha} \int\; dt\; 
E_j(p^{\alpha I}_t) \;\sigma(p^{\alpha I}_t,e)
\ee
Interestingly, if $E$ does not vary too much on the scale of a 
plaquette, then (\ref{2.32j}) actually reduces to (\ref{2.32k}) which is 
written directly in terms of the signed intersection number and 
plaquette fluxes. This will be useful later on when we compute 
expectation values.

\subsection{Comparison of Gauge Covariant Flux and Area Coherent States}
\label{s2.5}

Consider the case that the plaquttes are much smaller than 
the edges with respect to the three metric to be approximated by the 
coherent states and that the edges do not wiggle much on the scale of 
the plaquettes. Then for each $I$ the number of 
stacks that do not 
contain a vertex of $\gamma$ but still intersect $\gamma$ drastically 
outnumbers the number of stacks 
that do contain a vertex. Moreover, among the edge free stacks, the 
number of stacks 
that intersect only one 
edge completely outnumbers the ones that intersect more than one edge. 
Finally,
among those with single edge intersections, the number of stacks that 
intersect the respective edge 
once completely outnumbers the ones that do more than once. For this 
reason, in these cases the expressions (\ref{2.28}) and (\ref{2.29})
can be replaced with good approximation by simpler expressions of the 
form 
\be \label{2.33}
\lambda_s=\frac{\ell_P^2}{2a^2}
\sum_{e\in E(\gamma)}\; l^\gamma_e \; j_e(j_e+1)
\ee
and 
\be \label{2.34}
\lambda_s=\frac{\ell_P^2}{2a^2}
\sum_{e\in E(\gamma)}\; l^\gamma_e \; [n_e^j]^2
\ee
respectively where the length function $l^\gamma_e=g^\gamma_{ee}$ solves 
$l^\gamma_{e\circ 
e'}=l_e+l_{e'},\;l_{e^{-1}}=l_e$ in order that the complexifier 
has cylindrically consistent projections. 
This is the form of the heat kernel eigenvalue considered for the 
states in \cite{34}. As shown in \cite{21}, these eigenvalues  
cannot come from a known classical
complexifier so that the complexification map $A\mapsto Z$, without 
which the $Z$ label of the coherent state has no relation to the phase 
space point to be approximated, is unknown. When using 
the 
complexifier coming from a polyhedronal cell complex, a concrete 
relation between $Z$ and the phase space can be given for 
specific 
graphs, the above eigenvalues arise as we saw in section \ref{s2.3.1}
and as shown in \cite{21}.

Let us also check that the area complexification map $Z$ in 
(\ref{2.20}) and (\ref{2.21}) comes close to the gauge covariant flux 
one (\ref{2.16e}), at least on certain graphs. 
Let $\gamma$ be a graph dual to the cell complex $P$. Thus, for each 
edge $e$ there is a unique face $S_e$ which intersects $e$ in an 
interior point transversely such that $\sigma_{S_e\cap e}(S_e,e)=+1$
and no other face intersects $e$.
Then the gauge covariant flux complexification map at the 
level 
of the holonomies is given by \cite{10}
\be \label{2.35}
A(e)\mapsto g_e(Z):=Z_\gamma(e)=\exp(-i\tau_j E_j(S_e)/L^2)\; A(e) 
\ee
For $SU(2)$, $i\tau_j$ are the Pauli matrices while for 
$U(1)^3$ $i\tau_j=1,\;j=1,2,3$.  In contrast, the area complexification 
map
is given at the level of the holonomies by
\be \label{2.36}
A(e)\mapsto Z(e)={\cal P} \exp(\int_e Z^j \tau_j)
\ee
where $\cal P$ denotes path ordering where $Z_a^j$ is given in 
(\ref{2.20}) and (\ref{2.21}) respectively. Now for sufficiently 
``short'' edges we have $Z(e)\approx \exp(\int_e [Z-A]^j\tau_j) A(e)$
to leading order in the edge parameter length. 
If we assume that $E^a_j$ is slowly varying at the scale of the 
plaquettes then we have ${\rm Ar}(p^I_{\alpha_I(x) t_I(x)}) \approx
\sqrt{[E^a_j(x) n_a^I(x)]^2}$ so that 
(\ref{2.20}) is approximated by  
\be \label{2.37}
\int_e (Z^j-A^j)\approx -\frac{i}{a^2}\sum_I 
\int_0^1 \; \frac{\dot{e}^a(t) n_a^I(e(t))}{J_I(e(t))} \; 
n_b^I(e(t)) E^b_j(e(t))
\ee
where we have assumed that $e$ is the embedded interval $[0,1]$.
Now consider the case that the graph is in fact cubic and that the stack 
family and the graph are aligned in the following sense:\\
Suppose that we have an embedding
$X:\; \mathbb{R}^3 \to \sigma; s \mapsto X(s)$. For $\epsilon_{IJK}=1$ 
we define $X^I_t(u^1,u^2):=X(s^I=t,s^J=u^1,s^K=u^2)$. This defines 
linearly independent 
foliations $F^I$ with leaves $L_{It}=X^I_t(\mathbb{R}^2)$. 
The corresponding stack families are labelled by 
$\alpha:=(\alpha^1,\alpha^2)\in \mathbb{Z}^2$ and defined by 
$X^I_{\alpha t}:\; [0,1)^2 \to \sigma;\;X^I_{\alpha t}(u):=
X^I_t([\alpha^1+u^1]l,[\alpha^2+u^2]l)$ where $l>0$ is a 
certain parameter. The edges of the 
cubic graph are labelled by verticies $v=(v^1,v^2,v^3)\in \mathbb{Z}^3$ 
and directions $I$ and are defined for $\epsilon_{IJK}=1$ by 
$e_{v,I}:\;[0,1]\to \sigma; e_{v I}(t):=X(s^I=[v^I+t]\delta, 
s^J=v^J\delta, s^K=v^K\delta)$ where $\delta>0$ is another 
parameter. 

In this situation, (\ref{2.37}) can be further simplified to
\be \label{2.38}
\int_{e_{v I}} (Z^j-A^j)\approx 
-\frac{i}{a^2} \delta\;
\int_0^1 \;  n_b^I(e_{v I}(t)) E^b_j(e_{I v}(t))
\approx -\frac{i}{a^2} \delta E_j(p^I_v)
\ee
where $p^I_v$ is any plaquette in the stack in $I$ direction intersected 
by $e_{I v}$. 

Thus, for cubic 
graphs, which are the only ones considered so far in semiclassical 
calculations, we get a close match 
between (\ref{2.35}) and (\ref{2.38}) whenever the cubic graph and the 
stack families are aligned. However, there is still an 
important difference:\\ 
The parameter area $l^2$ of the plaquette $p^I_v$ in (\ref{2.38}) 
has no a priori relation to 
the parameter length $\delta$ of the edge $e_{v I}$ while the 
parameter area of the dual face $S_{e_{v I}}$  
in (\ref{2.35}) is of the order $\delta^2$. 
These considerations reveal that the individual plaquettes 
of the stacks cannot be interpreted as the faces of a dual graph
although roughly $[\delta/l]^2$ of them combine to a face. Hence the 
states considered in \cite{21} are genuinely different from those in 
\cite{10}.

This will 
turn out to be important:\\
We will see that in order to be able to perform practical calculations
for $SU(2)$ with off diagonal edge metrics, we need $l\ll \delta$ in 
order that the edge metric is close 
to diagonal for generic graphs. 
It turns out that if we use the same 
parameter $a$ in the label $Z$ of the state and for the classicality 
parameter $t=\ell_P^2/a^2$ then the expectation value of the volume 
turns out to be 
of the order of $(l/\delta)^3$ too small. Hence, there is a tension 
between the possibility to perform practical calculations and the 
correctness of the classical limit. The only analytical calculation 
possible with $l=\delta$ uses a graph which is aligned with the stacks 
and thus is necessarily cubic. While the result of that calculation 
results 
in the correct classical limit, this calculation is of limited interest 
because we saw already above that for this case the coherent states 
of \cite{21} reduce to those of \cite{10} for which we knew already that 
the classical limit is correct. 

However, the purpose of this paper is to test the semiclassical limit 
for graphs of non cubic topology. This can be done with the states of 
\cite{10} without limitation. With the states of \cite{21} this is 
possible if we 
redefine $Z_a^j \to A_a^j+\frac{a^2}{b^2}(Z_a^j-A_a^j)$ where $b\ll a$.
This rescaling is actually not in the spirit of the complexifier 
programme, but it repairs the semiclassical limit of {\it all operators} 
built from the fluxes.   
It will then turn out that for graphs that satisfy $l/\delta=b/a$ the 
correct classical limit results for $n=6$ only. As already mentioned 
in the introduction, one could rescale the label of the coherent state 
by a different amount in order to reach the correct semiclassical limit 
of the volume operator for one and only one $n\not=6$. However, that 
would destroy the correct semiclassical limit of other operators such as 
areas. Hence the rescaling by $(b/a)^2$ is harmless in the sense that 
it reproduces the semiclassical limit of all operators while 
$n-$dependent rescaling do not. 

Also with respect to the states of \cite{10} the value $n=6$ is 
singled out. The fact that the cut off states of \cite{21} have 
acceptable semiclassical behavior only when the corresponding cut off 
graph and the label of the coherent state satisfy certain restrictions 
imposed by the structure that defines the complexifier, in this case the 
size of the parquettes, is similar to the restrictions imposed on the 
by the polyhedronal cell complex complexifier \cite{10}, namely that the 
graph be dual to it.

\subsection{Justification for Replacing $SU(2)$ by $U(1)^3$}
\label{s2.6}

The considerations above have revealed that practically useful cut -- 
off states will be based on graphs which are much coarser than the 
parquets so that the edge metric is diagonal in very good approximation.
We will restrict to such graphs in the calculations that follow and find 
independent confirmation for that restriction as well in the form of the 
quality of the semiclassical approximation. Assuming exact diagonality
and thus suppressing the corrections coming from off -- diagonality 
which we will show to be small under the made coarseness assumptions,
the cut -- off states in fact factorizes
\be \label{2.39}
\psi_{Z, \gamma}=\prod_{e\in E(\gamma)} \; \psi_{Z,\gamma,e}
\ee
where for $SU(2)$
\be \label{2.40}
\psi_{Z,\gamma,e}(A)=\sum_{2j=0}^\infty\; (2j+1)\; 
e^{-\frac{t}{2} l^\gamma_e j(j+1)} \;
\chi_j(g_e A(e)^{-1})
\ee
while for $U(1)^3$
\be \label{2.41}
\psi_{Z,\gamma,e}(A)=\sum_{n\in \mathbb{Z}^3}\;  
e^{-\frac{t}{2} l^\gamma_e \sum_j (n^j)^2} \;
\chi_n(g_e A(e)^{-1})
\ee
Here $\chi_j$ and $\chi_n$ respectively denotes the character of the 
$j-$th and $n-$th irreducible 
representation of $SU(2)$ and $U(1)^3$ respectively. 

Under the made 
assumptions, the edge metrics 
$l^\gamma_e$ are identical for both groups because, while 
$l^{I \alpha \gamma}_N$ is defined for non negative integers $N$ only in 
the 
case of $SU(2)$ while for $U(1)^3$ all integers are allowed, for the 
graphs under consideration for each edge $e$ only either $N_e=+1$ or 
$N_e=-1$ leads to non vanishing $l^{I \alpha \gamma}_N$ so that these 
numbers in fact coincide and since we take the diagonal elements of the 
edge metric (\ref{2.32a}) both signs lead to the same $l^\gamma_e$.

Finally, if $(A_0, E_0)$ is the phase space point to be approximated 
and from which we calculate $Z=Z(A_0,A_0)$ via 
(\ref{2.20}) and (\ref{2.21}) then for $SU(2)$ we have, 
\be \label{2.42}
g_e\approx \exp(-i\tau_j P_0^j(e)) \; \exp(\tau_j \int_e A_0),\;\;
P_0^j(e)=\frac{1}{b^2} \sum_I \int_0^1\; dt \; \frac{\dot{e}^a(t) 
n_a^I(e(t))}{J_I(e(t))}\; [E^b_{0j}(e(t)) n_b^I(e(t))]\;    
\ee
while for $U(1)^3$ we have 
\be \label{2.43}
g_e=(g_e^j)_{j=1}^3,\;\; g_e^j=\exp(-P_0^j(e)+i\int_e A^j_0) 
\ee
where as before we have made the approximation 
\be \label{2.44}
\frac{{\rm Ar}(p^I_{\alpha_I(x) t_I(x)})}{\sqrt{[E^c_k(x) n_c^I(x)]^2}} 
\approx 1
\ee
which is valid if $E_0$ is slowly varying at the scale of the 
plaquettes.\\
\\
Thus, given $Z=Z(A_0,E_0)$, we have the following abstract situation 
under the 
made assumptions:\\
1.\\
For each edge $e$ there exist vectors $P_0^j(e), A_0^j(e)$ such that for 
$SU(2)$ we have $g_e\approx \exp(-i\tau_j P_0^j\tau_j) \exp(\tau_j 
A^j_0(e))\in SL(2,\mathbb{C})=SU(2)^{\mathbb{C}}$ while for $U(1)^3$ we 
have 
$g_e=(e^{-P^j_0(e)+iA_0^j(e)})_{j=1}^3\in 
(\mathbb{C}-\{0\})^3=(U(1)^3)^{\mathbb{C}}$.
2.\\
The coherent states adopt approximately product form 
$\psi_{Z,\gamma}\approx \prod_{e\in E(\gamma)} \; \psi_{g_e}$ where 
\be \label{2.45}
\psi_g(h)=\sum_{2j=0}^\infty\; (2j+1)\; e^{-t l^\gamma_e j(j+1)/2}\;
\chi_j(g h^{-1})
\ee
for $h\in SU(2)$ while
\be \label{2.46}
\psi_g(h)=\sum_{n\in \mathbb{Z}^3}\; e^{-t l^\gamma_e 
\sum_{j=1}^3 n_j^2}\;
\chi_n(g h^{-1})
\ee
for $h\in U(1)^3$.\\
\\
Now, as anticipated in the introduction, using the tools of 
semiclassical perturbation theory \cite{18} we are able to calculate 
to expectation value of the volume operator $V$ of LQG with respect to 
the 
correct $SU(2)$ coherent states in terms of the 
expectation value of of a certain operator $Q$, where $V=\root 4 
\of{Q}$,  
which we display 
explicitly in the next section and which is a sixth order polynomial
in the right invariant vector fields $X^j_e$ on $SU(2)$ where $X^j_e$ 
acts on $h_e$ in (\ref{2.45}). 
The crucial
observation, made in \cite{10}, is that if we simply replace the 
$SU(2)$ right invariant vector fields in $Q$ by $U(1)^3$ right 
invariant vector fields $X^j_e$ acting on $h_e$ in (\ref{2.46}) and if 
we replace the $SU(2)$ coherent states (\ref{2.45}) by the related 
$U(1)^3$ 
coherent states in (\ref{2.46}), then the remarkable fact is that {\it 
the expectation values of polynomials of right invariant vector 
fields actually 
coincide to zeroth order in $\hbar$}. By the same argument, this will 
be also true if we perform the right invariant vector field replacement 
already at the level of $V$ rather than $Q$. This observation was also key 
in the semiclassical analysis of \cite{19,20}. 

This feature is maybe not 
as surprising as it looks at first sight because, after all, the coherent 
states for both groups are to approximate the same phase space points. 
The underlying reason is that  
that the classical phase space of the $SU(2)$ theory (i.e. the range of 
fields and the 
symplectic 
structure) and of the fictive $U(1)^3$ theory actually coincide. It is 
only when 
we add the dynamics of the theory as for instance the Gauss constraint
that we see a difference. The Gauss law is taken into account in two 
ways, first by using 
the appropriate group coherent states, here $SU(2)$ or $U(1)^3$ 
respectively, which is dictated by the fact that the underlying 
holonomies take values in the appropriate group. Secondly, one can 
construct quantum Gauss constraint invariant coherent states 
\cite{10,38} by averaging over the gauge group action at the verticies.
Denote this group averaging map by $\eta$. Then, as shown in 
\cite{10,38}, we have that $<\eta(\psi_{Z,\gamma}, A 
\eta(\psi_{Z,\gamma})>$
and $\psi_{Z,\gamma}, A\psi_{Z,\gamma}>$ agree to zeroth order in 
$\hbar$ (notice that the Gauss invariant Hilbert space is an honest 
subspace of the kinematical Hilbert space so that the same inner 
product can be used) for any Gauss invariant
operator $A$ such as the volume operator because the overlap function 
between coherent states peaked at different phase space points is 
sharply peaked\footnote{In more detail we have 
\be \label{2.47}
\eta(\psi_{Z,\gamma})=\int_{G^{|V(\gamma)|}}\; \prod_{v\in 
V(\gamma)} 
\; d\mu_H(g_v)\; \alpha_g(\psi_{Z,\gamma})
\ee
where 
$\alpha_g(\psi_{Z,\gamma})(A)=\psi_{Z,\gamma}(\alpha_g(A))$ and 
$[\alpha_g(A)](e)=g(b(e)) A(e) g(f(e))^{-1}$ where $b(e)$ and $f(e)$ 
respectively denote beginning and final point of $e$ respectively.
Now due to gauge covariance of the coherent states we have 
$\alpha_g(\psi_{Z,\gamma})=\psi_{\alpha_{g^{-1}}(Z),\gamma}$ so that
the gauge invariant coherent state expectation value of a gauge 
invariant operator becomes (using the invariance properties of the 
Haar measure)
\be \label{2.48}
\frac{<\eta(\psi_{Z,\gamma}), A\eta(\psi_{Z,\gamma})>}
{||\eta(\psi_{Z,\gamma})||^2}
=\frac{  \int_{G^{|V(\gamma)|}}\; \prod_{v\in V(\gamma)} \; d\mu_H(g_v)\;
<\psi_{\alpha_g(Z),\gamma},A\psi_{Z,\gamma}>}
{\int_{G^{|V(\gamma)|}}\; \prod_{v\in V(\gamma)} \; d\mu_H(g_v)\;
<\psi_{\alpha_g(Z),\gamma},\psi_{Z,\gamma}>}
\ee
Now from \cite{10} we know for gauge invariant polynomials $A$ in 
right invariant vector fields that the  
peakedness property 
\be \label{2.49}
<\psi_{Z',\gamma},A \psi_{Z,\gamma}>=
\frac{<\psi_{Z,\gamma}, A \psi_{Z,\gamma}>}{||\psi_{Z,\gamma}||^2}
\;<\psi_{Z',\gamma},\psi_{Z,\gamma}>\;[1+O(\hbar)]
\ee
holds. Now the claim is immediate.}. This justifies the use 
of the kinematical states employed in this paper.\\
\\
To summaries:\\
Using kinematical $U(1)^3$ coherent states is a convenient approximation 
for actual $SU(2)$ coherent state expectation value calculations for 
Gauss invariant operators if one 
is only interested in the zeroth order in $\hbar$. At non vanishing 
orders in $\hbar$ there will be differences but we are not interested in 
them for the purposes of this paper. One may wonder whether the argument 
made above, namely using kinematical rather than Gauss invariant 
coherent states also survives when considering the spatial 
diffeomorphism constraint. This issue, currently 
under investigation, is more complicated in part because it is not 
completely obvious which distributional extension of the classical 
diffeomorphism group one should use \cite{39}. For the purposes of this 
paper this is of no concern because we are looking at the {\it 
local} volume operator which is not spatially diffeomorphism 
invariant so that expectation value calculations with respect to 
spatially diffeomorphism invariant coherent states are meaningless. It 
is the local volume which enters the Hamiltonian and Master constraint 
and verifying the semiclassical limit of those only makes sense at the
kinematical Hilbert space level (one cannot check the correct classical 
limit of a constraint on its kernel). Once this limit is verified, one 
has 
confidence that the physical Hilbert space defined by the Hamiltonian 
constraint is correct.

\section{Regular Simplicial, Cubical and Octahedronal Cell Complexes}
\label{s3}

In order to perform the calculations in our companion paper \cite{30b} 
for the 
coherent states of \cite{21}, we need the 
specific embedding of the $n=4,6,8$ valent graph relative to the stack
families. This can easiest be done by starting from regular dual 
simplicial 
(tetrahedronal), cubical and octahedronal partitions of the three 
manifold $\sigma$. For the coherent states of \cite{10} this section
is not needed except that it shows the existence of (regular) polyhedral
cell complexes dual to $n=4,6,8$ valent graphs such that all cells of 
that complex are platonic solid bodies, i.e. tetrahedral, cubes and 
octahedra respectively.

\begin{figure}[hbt]
\begin{center}
 \includegraphics[scale=0.3]{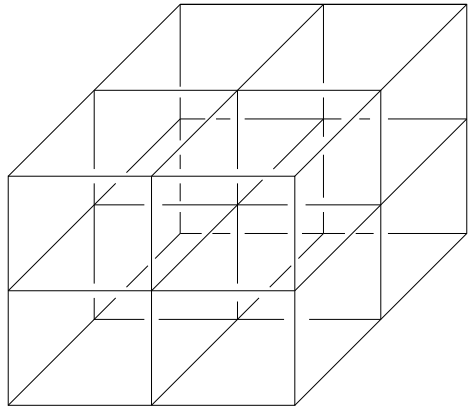}
\caption{Cubic cell decomposition.}
 \label{fig2}
\end{center}
\end{figure}

\begin{figure}[hbt] 
\begin{center}
 \includegraphics[scale=0.3]{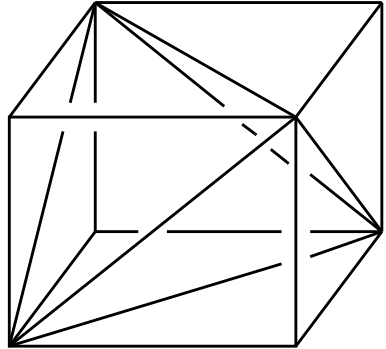}
\caption{Type A triangulation of a cube.}
\label{fig3}
\end{center}
 \end{figure}

\begin{figure}[hbt] 
\begin{center}
 \includegraphics[scale=0.3]{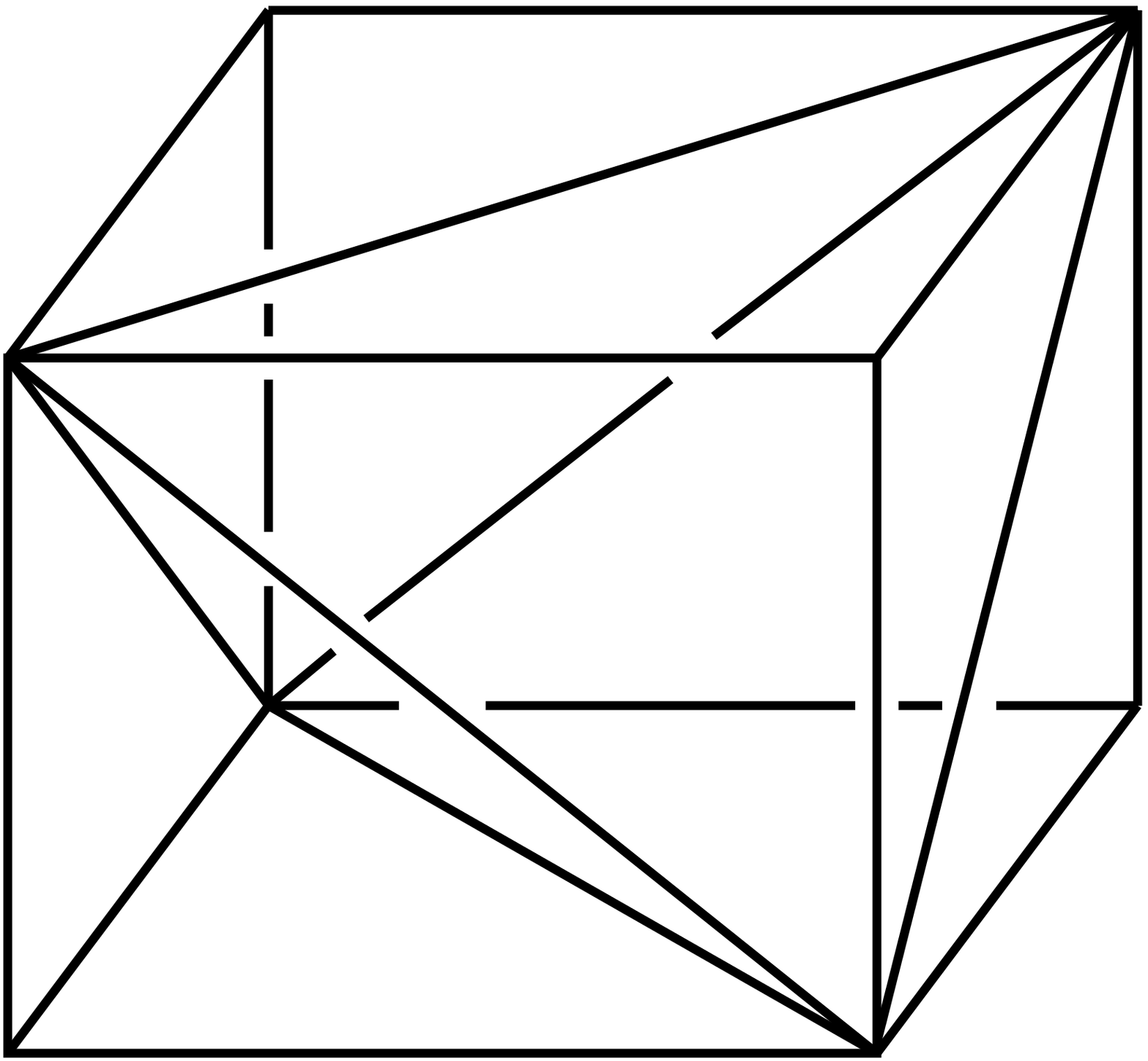}
\caption{Type B triangulation of a cube.}
\label{fig4}
\end{center}
 \end{figure}

\begin{figure}[hbt] 
\begin{center}
 \includegraphics[scale=0.3]{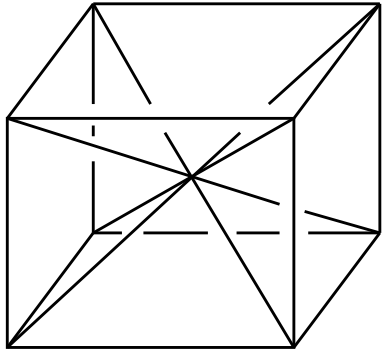}
\caption{Decomposition of a cube into six pyramids.}
\label{fig5}
\end{center} \end{figure}

In fact, it is possible to define such partitions all from refinements 
of cubical decompositions such as sketched in figure \ref{fig2}.
We perform the analysis 
for each chart 
$X:\;\mathbb{R}^3\to \sigma$ separately and use the Euclidean metric on 
$\mathbb{R}^3$ in the following definitions. 
\begin{Definition} \label{def3.1} ~\\
i.\\
A cubical partition of $\mathbb{R}^3$ is defined by the cubes 
$c_n,\;n\in \mathbb{Z}^3$ where 
\be \label{3.1}
c_n=\{s\;\in \mathbb{R}^3;\; s^I=n^I+t^I,\;I=1,2,3\}
\ee
The boundary faces (squares) of $c_n$ are taken with outward 
orientation.\\
ii.\\
A simplicial partition of $\mathbb{R}^3$ subordinate to a cubical one 
is defined as follows:\\
First draw in $c_{(0,0,0)}$ diagonals on the boundary squares such that 
the 
diagonals on opposite squares are orthogonal. Specifically, in the face 
defined by $s^I=0;\; s^J,s^K\in [0,1]^2;\; \epsilon_{IJK}=1$ the 
diagonal is the line $t\mapsto (s^I=0,s^J=t,s^K=t),\;t\in [0,1]$ while    
in the face 
defined by $s^I=1;\; s^J,s^K\in [0,1]^2;\; \epsilon_{IJK}=1$ the 
diagonal is the line $t\mapsto (s^I=1,s^J=t,s^K=1-t),\;t\in [0,1]$.\\
Now continue this pattern of orthogonal diagonals in opposite faces to 
the six cubes adjacent to $c_0$ where common faces have the same 
diagonal. This also defines the remaining four diagonals in those six 
cubes by connecting the endpoints of the already present two diagonals. \\
Finally continue this process for all cubes.\\
The face diagonals define altogether five tetrahedra that partition each 
cube. We will take their boundary triangles with outgoing orientation.\\
iii.\\
An octahedronal partition of $\mathbb{R}^3$ subordinate to a cubical
one is defined as follows:\\
For each cube draw the unique four space diagonals. Specifically in 
$c_{(0,0,0)}$ these are the 
lines $t\mapsto (t,t,t),\;(t,t,1-t),\;(t,1-t,t),\;(1-t,t,t);\;\;
t\in [0,1]$. These partition each cube into six pyramids with common tip
in the barycenter of the cube and with the six faces of the cube as 
their bases. Now glue two pyramids in adjacent cubes 
along their common base. Obviously, two glued pyramids define an 
octahedron which we take with outgoing orientation. 
\end{Definition}  
The basic building blocks of the tetrahedronal and octahedronal 
decompositions are displayed in figures \ref{fig3}, \ref{fig4} and 
\ref{fig5} respectively.
When gluing the bases of the pyramids along the faces of the original 
cubes one obtains an octahedronal decomposition as displayed in figure 
\ref{s6}.
\begin{figure}[hbt] 
\begin{center}
 \includegraphics[scale=0.3]{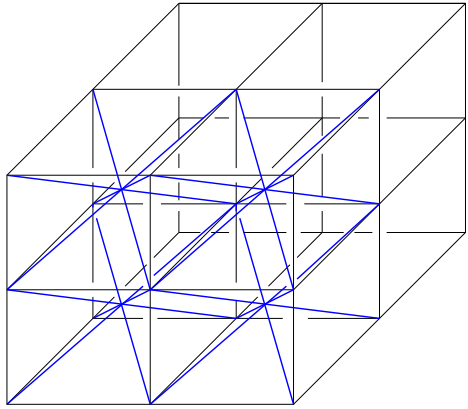}
\caption{Octahedronal decomposition.}
\label{fig6} 
\end{center} 
\end{figure}
It is maybe not completely obvious that the 
drawing of the diagonals that are to define the tetrahedra is a 
consistent and unique prescription. To 
see this, we use the checkerboard visualization displayed in 
figure{fig7}: 
\begin{figure}[hbt] 
\begin{center}
 \includegraphics[scale=0.3]{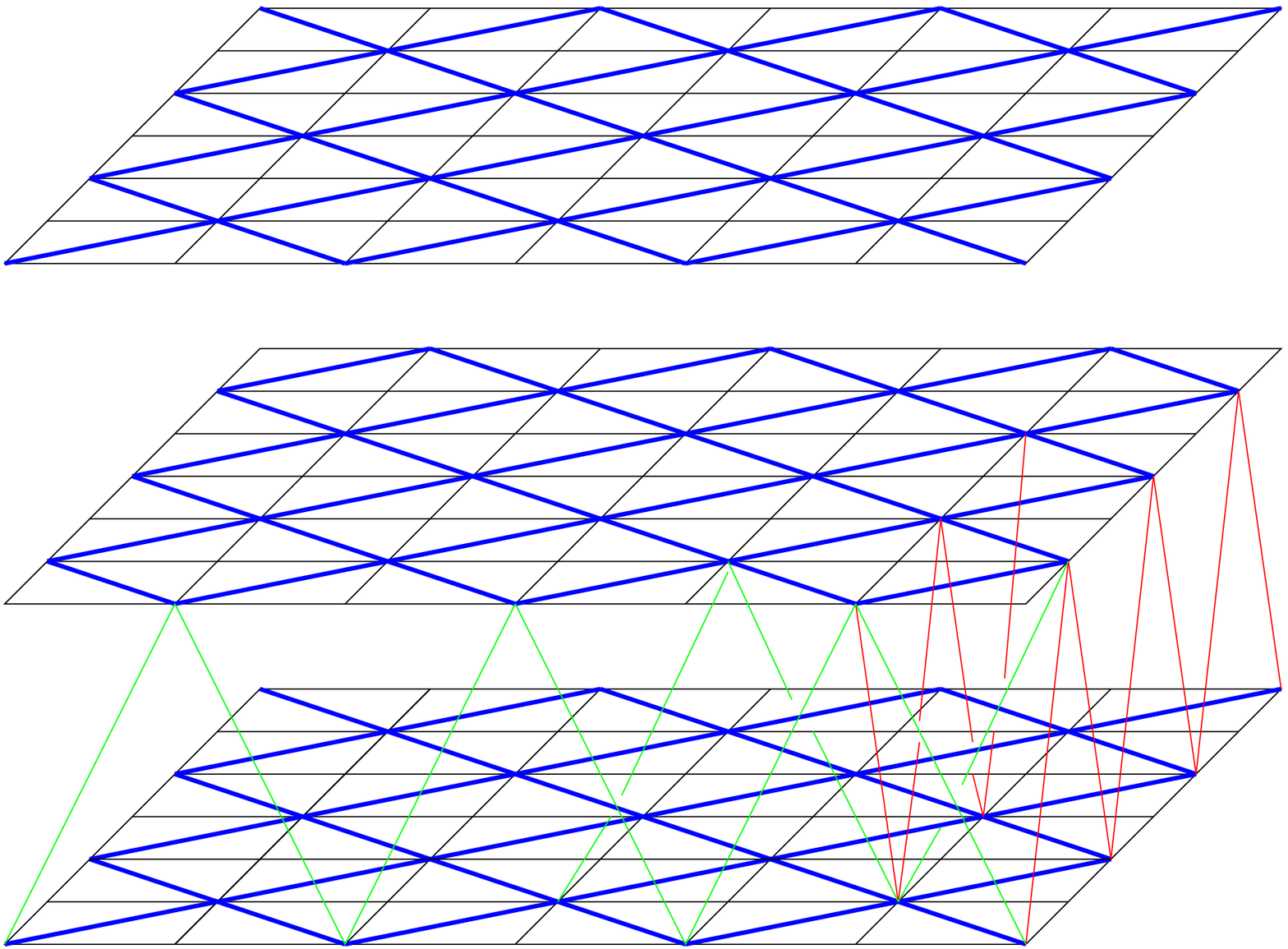}
\caption{Checkerboard visualization of the triangulation.}
\label{fig7}
\end{center} 
\end{figure}
First draw all 
plaquettes in the $s^3=n\in\mathbb{Z}$ layers. 
Now take the $n=0$ layer and draw the diagonal for the plaquette in that 
layer that belongs to $c_{(0,0,0)}$ as prescribed in the definition.
Define that plaquette as ``black''. Now turn the $n=0$ layer into a 
checkerboard in the unique way consisting of black and white plaquettes.
The other layers $n\not=0$ are also turned uniquely into  
checkerboards by asking that checkerboards in adjacent layers are  
complementary, i.e. if the plaquette $(n^1,n^2,n^3)$ is white (black)
then the plaquette $(n^1,n^2,n^3\pm 1)$ is black (white). 
Draw diagonals in plaquettes of opposite colour orthogonally to each 
other. This defines face diagonals in the $s^3=n$const. layers. 
These have the property that they form squares in each layer which 
lie at an angle of $\pi/4$ relative to the plaquettes and which are such 
that only every second plaquette corner is a vertex of these squares.
We will refer to such corners that are verticies as ``used''. 
It is easy to see that in adjacent layers, used plaquette corners lie 
above unused ones.  Now draw the 
remaining face diagonals in the $s^1,s^2=n=$const. layers by connecting 
the used corners in adjacent layers using the appropriate diagonals of 
the cubes. This results in the triangulation depicted in figure 
\ref{fig8}.
\begin{figure}[hbt] 
\begin{center} 
\includegraphics[scale=0.3]{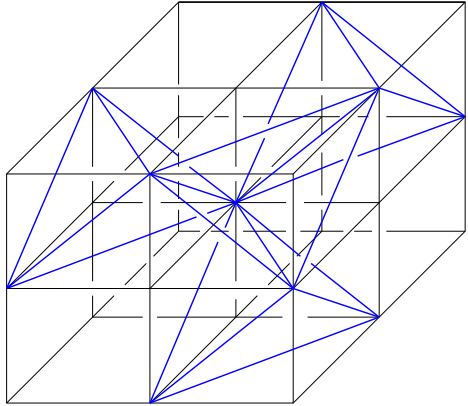}
\caption{Triangulation.}
\label{fig8}
\end{center} \end{figure}
We now define the graphs dual to these particular polyhedronal 
decompositions.
\begin{Definition} \label{def3.2} ~\\
The graph in $\mathbb{R}^3$ dual to the above simplicial, cubical and 
octahedronal cell complexes is obtained by connecting the barycentres of 
adjacent tetrahedra, cubes and octahedra respectively by straight lines 
through their common triangles, squares and triangles respectively. 
Here the barycenter of a region $R\subset \mathbb{R}$ is defined as 
usual by 
\be \label{3.3}
B(R)=\frac{\int_R\; d^3s\; (s^1,s^2,s^3)}{\int_R\; d^3s}
\ee
\end{Definition}
The advantage of the explicit definition of the cell complex is that we 
can explicitly label the edges and verticies of the dual graph. This is 
of course only feasible for sufficiently regular graphs, otherwise we 
run into difficult bookkeeping problems. 
\begin{itemize}
\item[1.] {\it Cubical Graph}\\
The barycentres of the cubes 
$c_n$ are evidently the points 
$v_n:=(n^1+\frac{1}{2},n^2+\frac{1}{2},n^3+\frac{1}{2})$ which form 
the
verticies of the dual graph. The edges $e_{n,I},\;I=1,2,3$ which connect
the verticies with labels $n$ and $n+b_I$ respectively, where $b_I$ is 
the standard unit vector $(b_I)^J=\delta_I^J$, have the explicit 
parametrization 
$e_{n,I}(t)=v_n+tb_I,\; t\in [0,1]$. The other three edges adjacent to 
$v_n$ are ingoing and are given by $e_{n-b_I,I}$.
These 
edges form the 1 skeleton of 
another cubical cell complex which is just shifted by the vector 
$(\frac{1}{2},\frac{1}{2},\frac{1}{2})$ from the original one.
\item[2.] {\it Tetrahedronal graph}\\
The tetrahedronal graph is the most complicated one because there are two 
different types of simplicial decompositions of a cube into five 
tetrahedra. Type A. corresponds to the case that the verticies of the 
internal tetrahedron within a standard unit cube are given by 
$(0,0,0),\;(1,1,0),\;(1,0,1),\;(0,1,1)$ while type B. has verticies at
$(1,0,0),\;(0,1,0),\;(0,0,1),\;(1,1,1)$. These types alternate in adjacent 
cubes as we move in any of the three coordinate directions. 
Hence, by defining the cube $c_0$ to be of type A., 
the triangulation is completely completely specified. Indeed, the type of 
$c_n$ is A. if $n^1+n^2+n^3$ is even and of type B. otherwise.  

To determine the dual graph, we first discuss the barycentres of the 
tetrahedra for the two 
types separately for 
a standard unit cube as well as the edges of the dual graph that lie 
within it. The verticies of and the edges 
in $c_n$ follow then by translation 
by $n=n^I b_I$. Notice that 
a tetrahedron $T$ based at $v$ and spanned by vectors $e_I$, that is,
$T=\{v+t^I e_I;\; 0\le t^I\le 1;\; t^1+t^2+t^3\le 1\}$, has 
barycenter at $B(T)=v+\frac{1}{4}(e_1+e_2+3_3)$.
\begin{itemize}
\item[A.] {\it Type A.}\\
The barycenter of the interior tetrahedron coincides with the barycenter 
$v_0:=\frac{1}{2}(1,1,1)$ of the cube. The barycentres of the remaining 
four 
exterior tetrahedra 
based at verticies\\  $(1,0,0),\;(0,1,0),\;(0,0,1),\;(1,1,1)$
respectively are at
$v^A_1:=\frac{1}{4}(3,1,1),\;v^A_2=\frac{1}{4}(1,3,1),\;
v^A_3=\frac{1}{4}(1,1,3),\;v^A_4=\frac{1}{4}(3,3,3)$ respectively. 
Accordingly, the dual edges within the cube are 
$e^A_\alpha=v^A_\alpha-v_0,\;\alpha=1,2,3,4$.
\item[B.] {\it Type B.}\\
The barycenter of the interior tetrahedron coincides with the barycenter 
$v_0:=\frac{1}{2}(1,1,1)$ of the cube. The barycentres of the remaining 
four 
exterior tetrahedra 
based at verticies  
$(0,0,0),\;(1,1,0),\;(1,0,1),\;(0,1,1)$\\ 
respectively are at
$v^B_4:=\frac{1}{4}(1,1,1),\;v^B_3=\frac{1}{4}(3,3,1),\;
v^B_2=\frac{1}{4}(3,1,3),\;v^B_1=\frac{1}{4}(1,3,3)$ respectively. 
Accordingly, the dual edges within the cube are 
$e^B_\alpha=v^B_\alpha-v_0,\;\alpha=1,2,3,4$.
\end{itemize}
It remains to describe the dual edges that result from gluing the faces of 
the exterior tetrahedra of adjacent cubes. But this is simple because each 
of the exterior tetrahedra 
within a cube has three triangles as faces which lie in the three 
coordinate planes, hence the gluing is between those triangles which 
result from drawing the respective face diagonal within a boundary square 
of a cube. Hence, each cube has twelve edges perpendicular to the twelve 
boundary triangles of the exterior tetrahedra which are adjacent to the 
four barycentres of those exterior tetrahedra. Altogether we can identify 
six possible gluings, namely either going from type A. to type B. when 
moving along the positive $I$ direction and gluing along the  
$s^I=$const. plane or going from type B. to type A. when 
moving along the positive $I$ direction and gluing along the  
$s^I=$const. plane. As one may check, the type A. to type B. gluing
in $I$ direction corresponds to two dual edges running from $v_I^A$
to a $v_4^B$ and from $v_4^A$ to $v_I^B$ types of verticies respectively. 
Likewise,
the type B. to type A. gluing
in $I$ direction corresponds to two dual edges running from $v_J^A$
to $v_K^B$ and from $v_K^A$ to $v_J^B$ types of verticies respectively 
where $\epsilon_{IJK}$. In all cases, these $I$ direction edges have 
coordinate length $\frac{1}{2}$ as one may easily calculate.

Altogether, we can now easily describe the dual lattice as follows:\\
The verticies are labelled $v_{n,\alpha},\;\alpha=0,1,2,3,4$ with 
$v_{n,0}=n+v_0$ and $v_{n,\alpha}=n+v^A_\alpha,\;\alpha=1,2,3,4$ if 
$n^1+n^2+n^3$ is even while $v_{n,\alpha}=n+B^A_\alpha,\;\alpha=1,2,3,4$ 
if $n^1+n^2+n^3$ is odd. The edges are labelled by 
$e_{n,\alpha},\;\alpha=1,2,3,4$ and $e_{n,I,j},\;I=1,2,3,\;j=1,2$
where $e_{n,\alpha}(t)=n+v_0+t(v^A_\alpha-v_0$ if $n^1+n^2+n^3$ is even,  
$e_{n,\alpha}(t)=n+v_0+t(v^B_\alpha-v_0)$ if $n^1+n^2+n^3$ is odd,  
$e_{n,I,1}(t)=n+v_I^A+\frac{t}{2}b_I$ and    
$e_{n,I,2}(t)=n+v_4^A+\frac{t}{2}b_I$ if $n^1+n^2+n^3$ is even
and finally    
$e_{n,I,1}(t)=n+v_J^A+\frac{t}{2}b_I$ and    
$e_{n,I,2}(t)=n+v_K^A+\frac{t}{2}b_I$ if $n^1+n^2+n^3$ is odd
where $\epsilon_{IJK}=1$. 
\item[3.] {\it Octahedronal Graph}\\
Each cube contains six pyramids or halves of 
the octahedra. Therefore the barycenter of an octahedron coincides with 
the barycenter of the common boundary face of the two cubes that contain 
it. It follows that the octahedra may be labelled by $o_{n,I}$ 
corresponding to the verticies $v_{n,I}=n+\frac{1}{2} b_J+\frac{1}{2} 
b_K;\; \epsilon_{IJK}=1$ which define its barycenter. Such an octahedron
has the property that it has a common base of two pyramid halves which 
lies in the $s^I=$const. plane.   
For the vertex $v_{n,I}$ we define four edges 
$e_{n,I,J,\sigma},\;J\not=I;\;\sigma=\pm$ outgoing from it through the 
explicit 
parametrization
$e_{n,I,j}(t):=v_{n,I}+\frac{t}{2}(b_I+\sigma b_J)$ which connects 
the verticies $v_{n,I}$ and $v_{n+\frac{1}{2}(1+\sigma) b_J,J}$. 
Notice that these edges lie in the $(I,J)$ or $(I,K)$ plane but there 
are no edges in the $(J,K)$ plane adjacent to $v_{n,I}$.
The other four edges adjacent to $v_{n,I}$ have ingoing orientation.

As an aside, notice that the 1-skeleton of an octahedral cell complex as 
defined 
above is an eight valent graph after removing the edges of the original 
cubes. 
\end{itemize}
The basic building blocks of the dual graphs are displayed in figures
\ref{fig9}, \ref{fig10}, \ref{fig11} and \ref{fig12} respectively.
\begin{figure}[hbt] 
\begin{center} 
\includegraphics[scale=0.3]{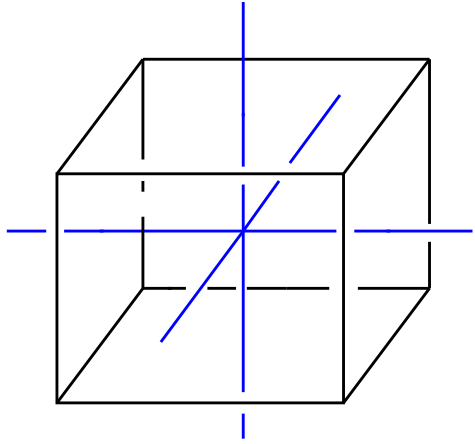}
\caption{Cube and dual six vallent graph.}
\label{fig9}
\end{center} \end{figure}
\begin{figure}[hbt] 
\begin{center}
 \includegraphics[scale=0.3]{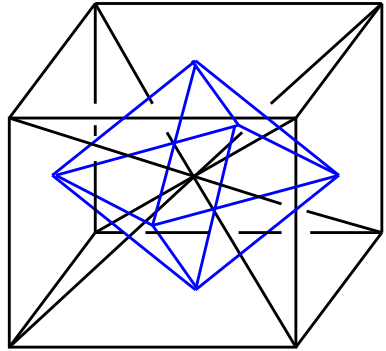}
\caption{Octahedron and dual eight valent graph.}
\label{fig10}
\end{center} \end{figure}
\begin{figure}[hbt] 
\begin{center} \includegraphics[scale=0.3]{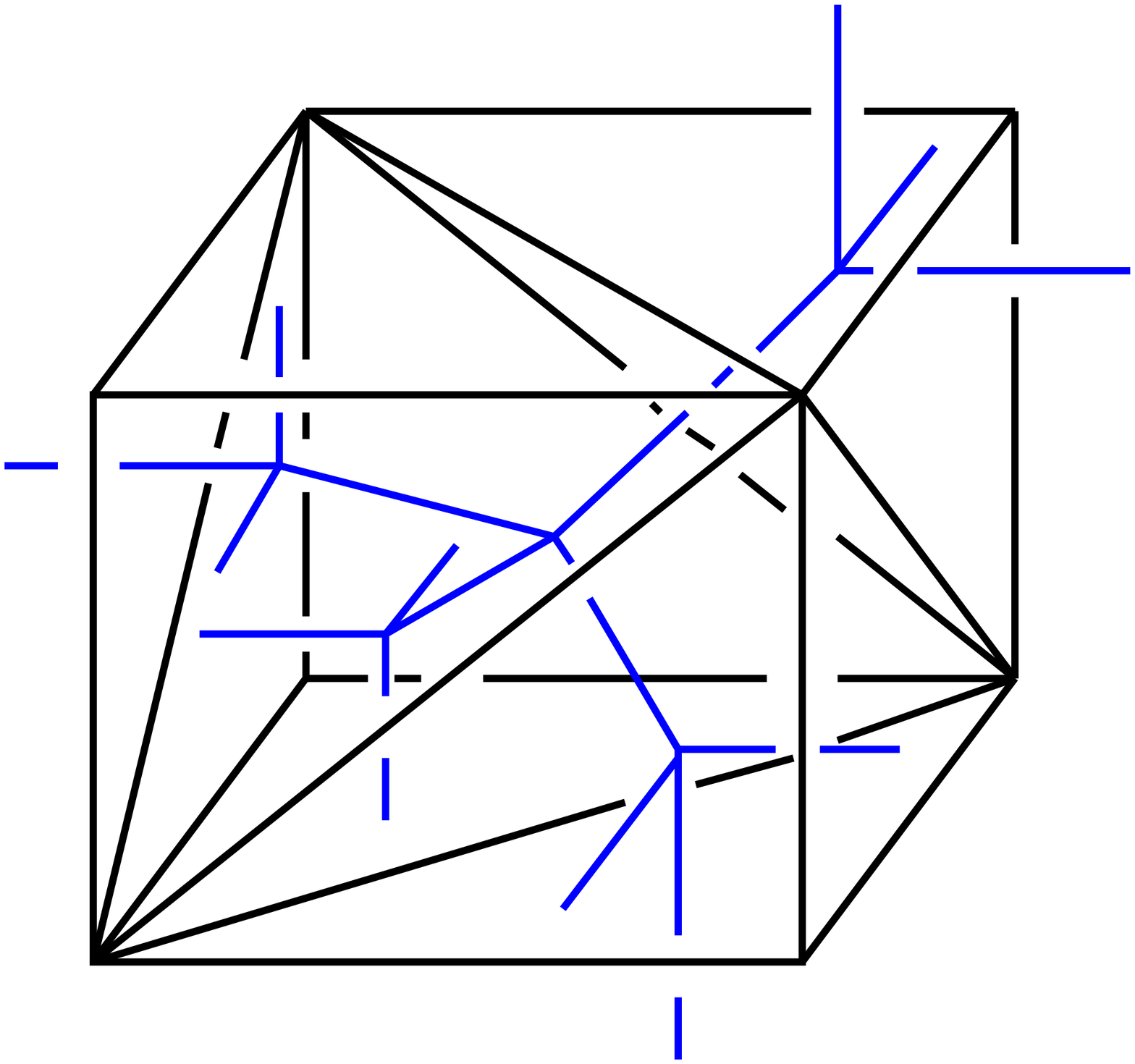}
\caption{Type A triangulation of a cube and dual four valent graph.}
\label{fig11}
\end{center} \end{figure}
\begin{figure}[hbt] 
\begin{center} 
\includegraphics[scale=0.3]{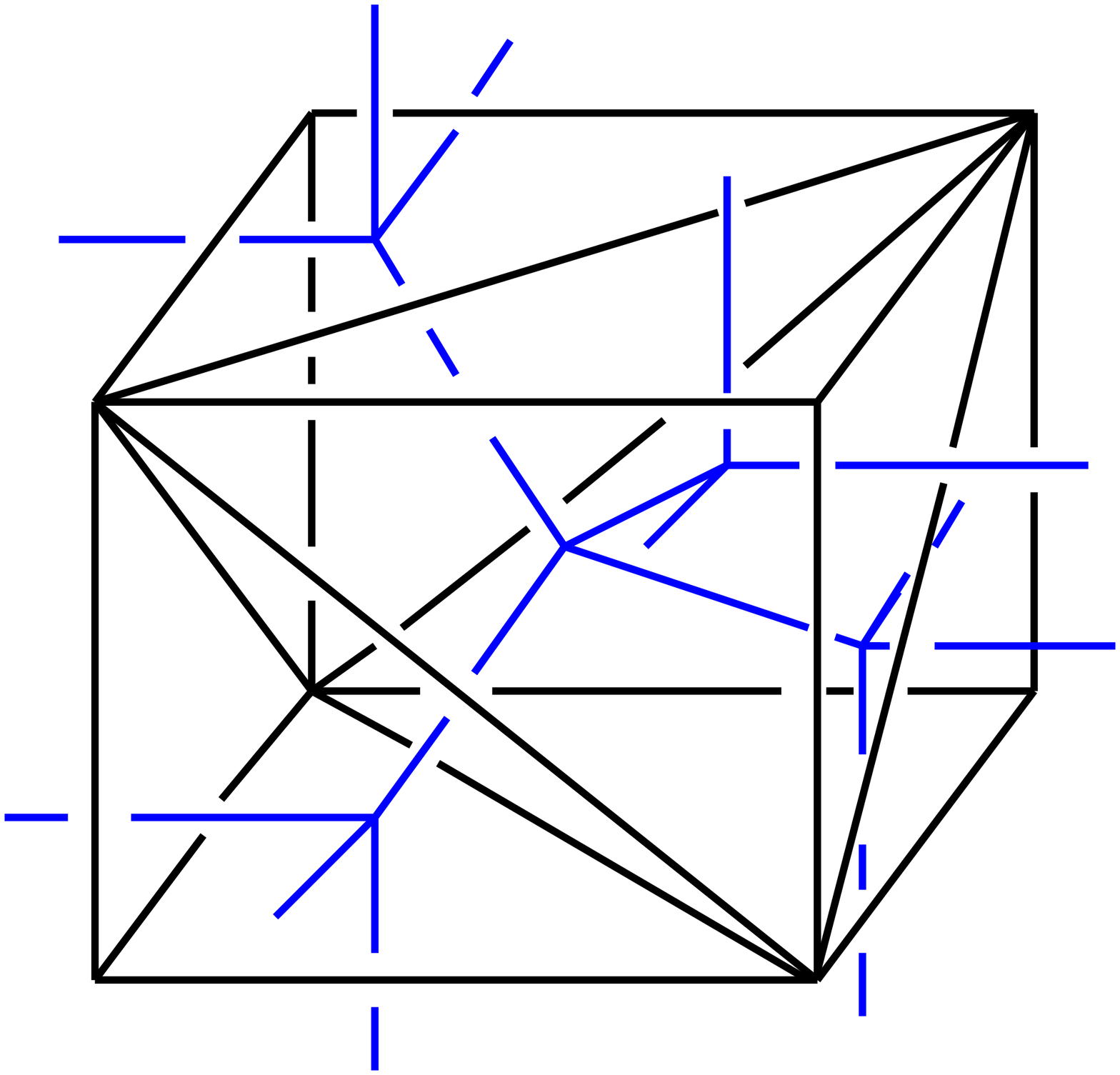}
\caption{Type B triangulation of a cube and dual four valent graph.}
\label{fig12}
\end{center} \end{figure}
The connection of the tetrahedronal lattice with the diamond lattice is as 
follows:\\
For each cube of type A. or B. respectively, keep the interior 
tetrahedron. Now move the barycentres of the remaining exterior tetrahedra 
into that corner of the cube which is also a corner of the tetrahedron 
under consideration. In this process, the edges dual to the faces of the 
interior tetrahedron become halves of the spatial diagonals of the cube.
Finally drop all the other edges which were running between the 
barycentres of the exterior tetrahedra. The result is a diamond lattice. 
Its basic building blocks are depicted in figures \ref{fig13} and 
\ref{fig14} respectively.
\begin{figure}[hbt] 
\begin{center}
 \includegraphics[scale=0.3]{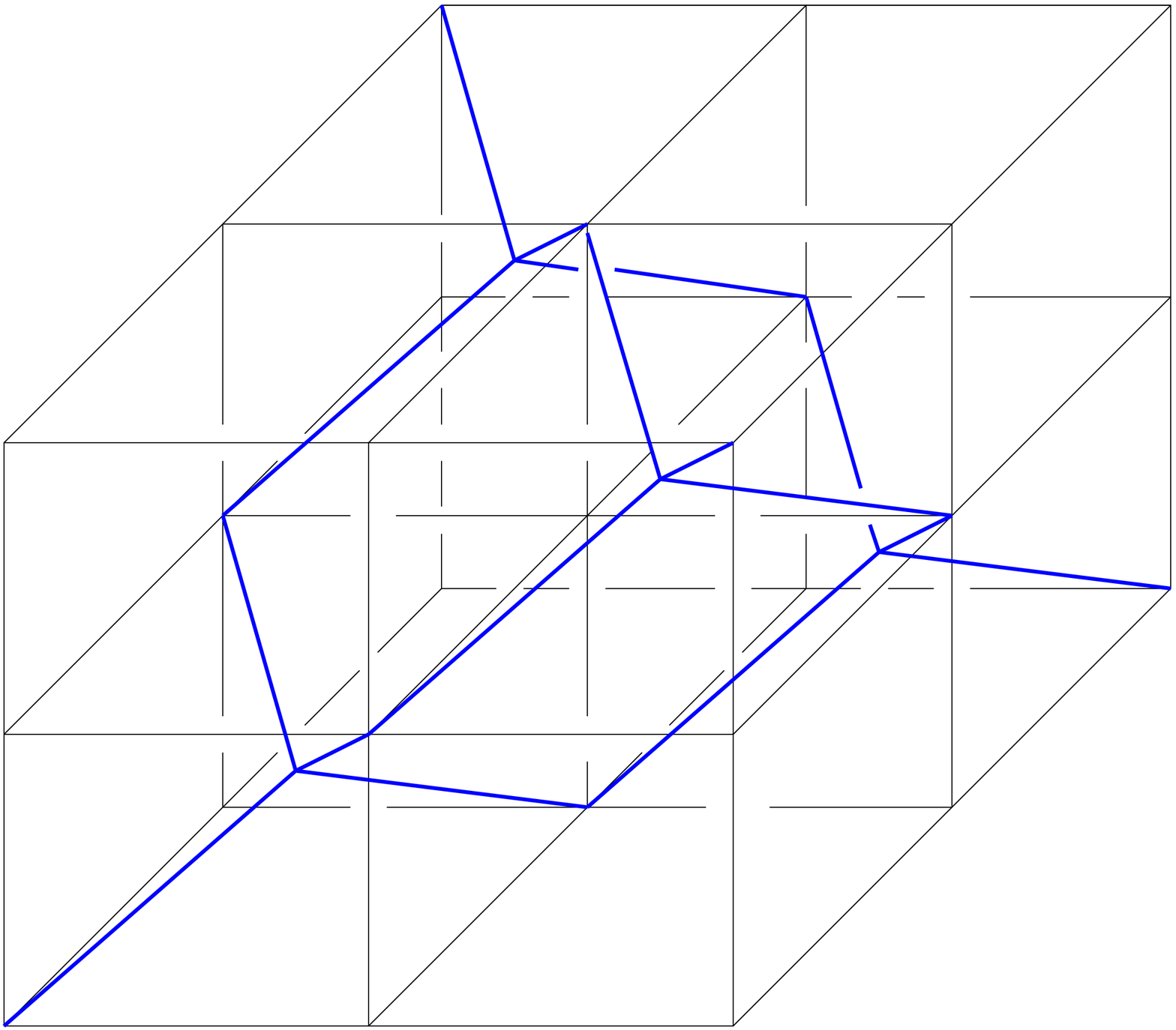}
\caption{Type A diamond cell with occupied lower, left front cube.}
\label{fig13}
\end{center} \end{figure}
\begin{figure}[hbt] 
\begin{center} 
\includegraphics[scale=0.3]{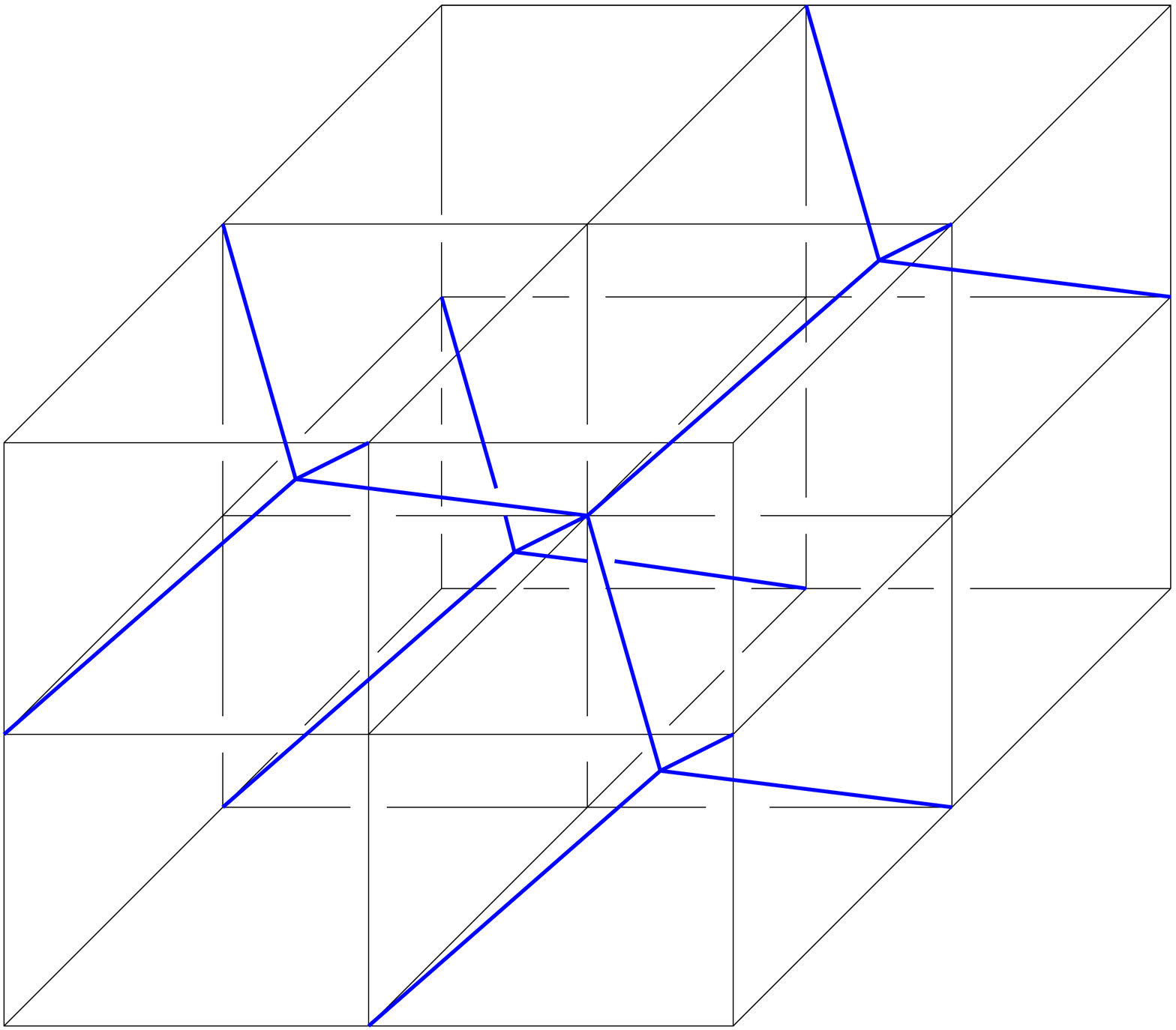}
\caption{Type B diamond cell with unoccupied lower, left front cube.}
\label{fig14}
\end{center} \end{figure}
It is also four valent, however, it does not have a piecewise 
linear polyhedronal complex 
dual to it (i.e. whose faces (which are subsets of linear planes) are in 
one to one correspondence with the 
edges). It does have a cell complex dual to it if one gives up 
piecewise linearity by suitably rounding off corners but that is 
inconvenient to describe analytically. On the other hand, the natural 
polyhedronal complex consisting of the 
interior 
tetrahedra of the original cubes with the cubes deleted consists 
of those tetrahedra as well octahedra which surround half of the corners 
of the original cubes. Only half of the triangle faces of those 
octahedra are penetrated by the edges of the diamond lattice. 
The building of this semi dual polyhedronal cell complex consisting of 
tetrahedra and octahedra respectively is visualized in figures 
\ref{fig15}, \ref{fig16}, \ref{fig17}, \ref{fig18} and \ref{fig19} 
respectively. 
\begin{figure}[hbt] 
\begin{center}
\includegraphics[scale=0.3]{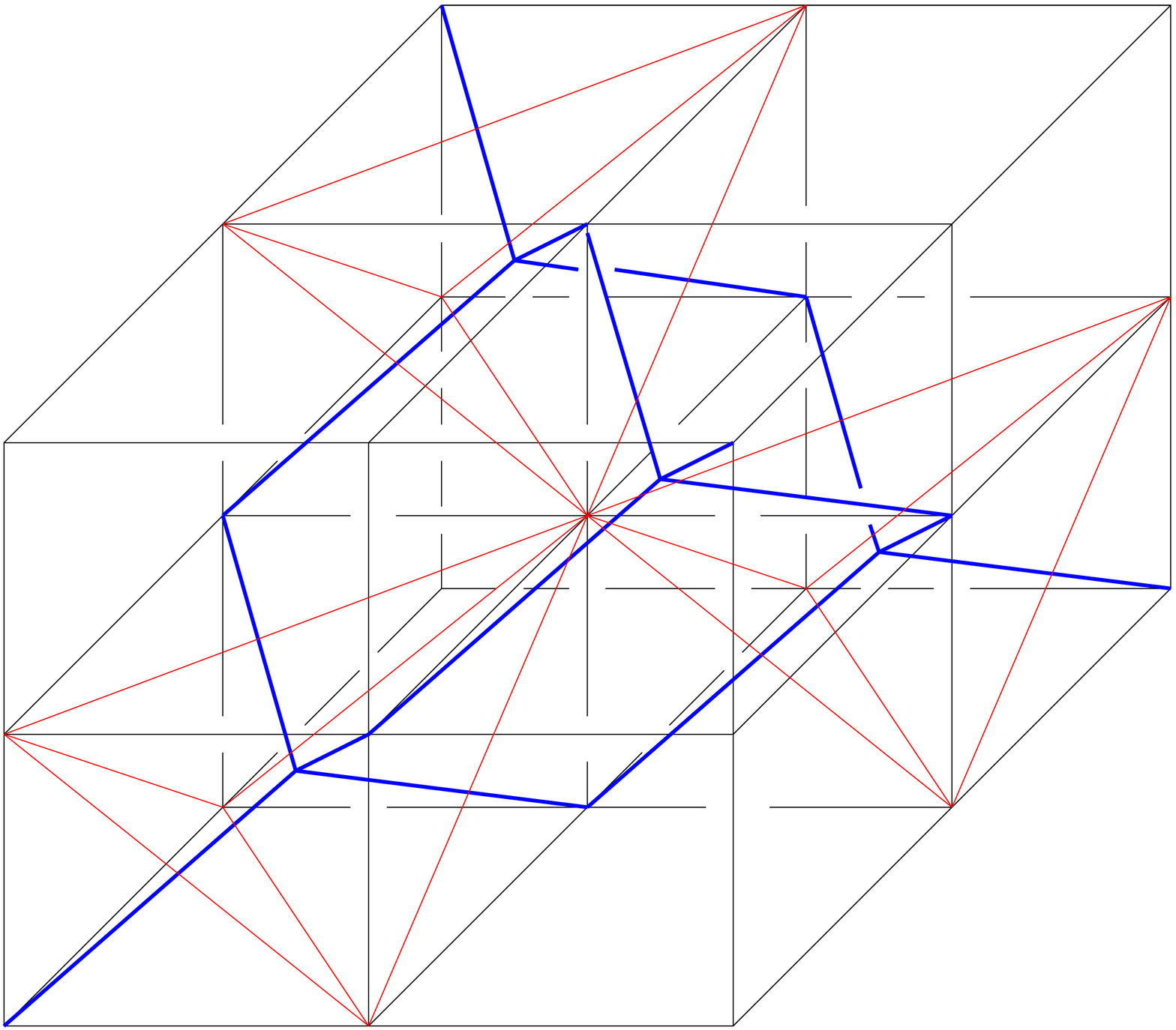}
\caption{Dual diamond cell of type A.}
\label{fig15}
\end{center} \end{figure}
\begin{figure}[hbt] 
\begin{center} \includegraphics[scale=0.3]{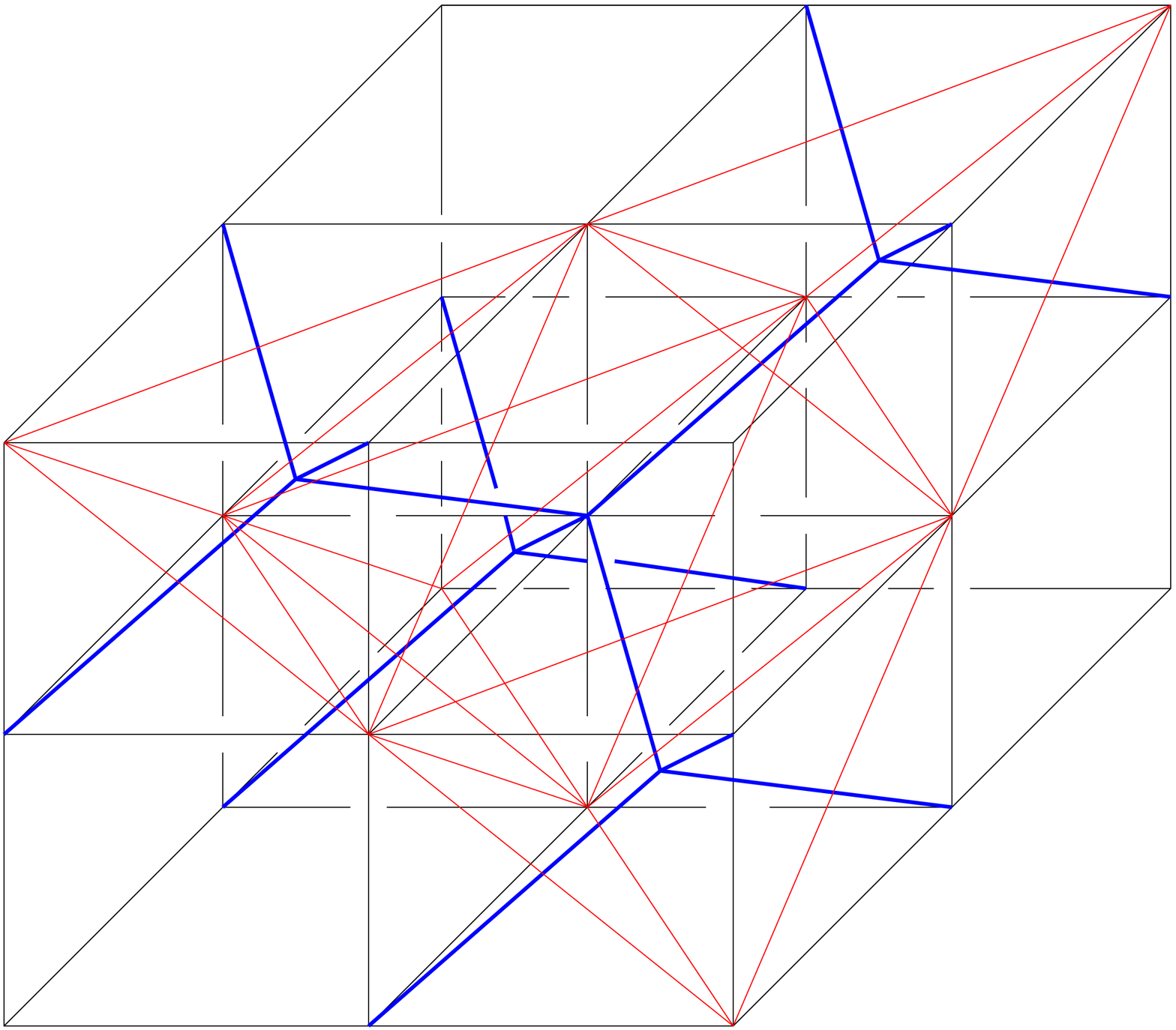}
\caption{Dual diamond cell of type B.}
\label{fig16}
\end{center} \end{figure}
\begin{figure}[hbt] 
\begin{center} \includegraphics[scale=0.3]{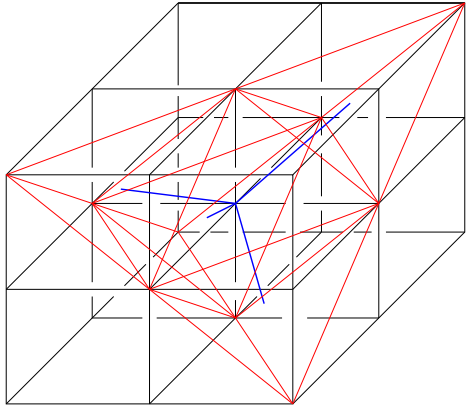}
\caption{Dual diamond cell of type B with only the central four valent 
vertex left.}
\label{fig17}
\end{center} \end{figure}
\begin{figure}[hbt] 
\begin{center} \includegraphics[scale=0.3]{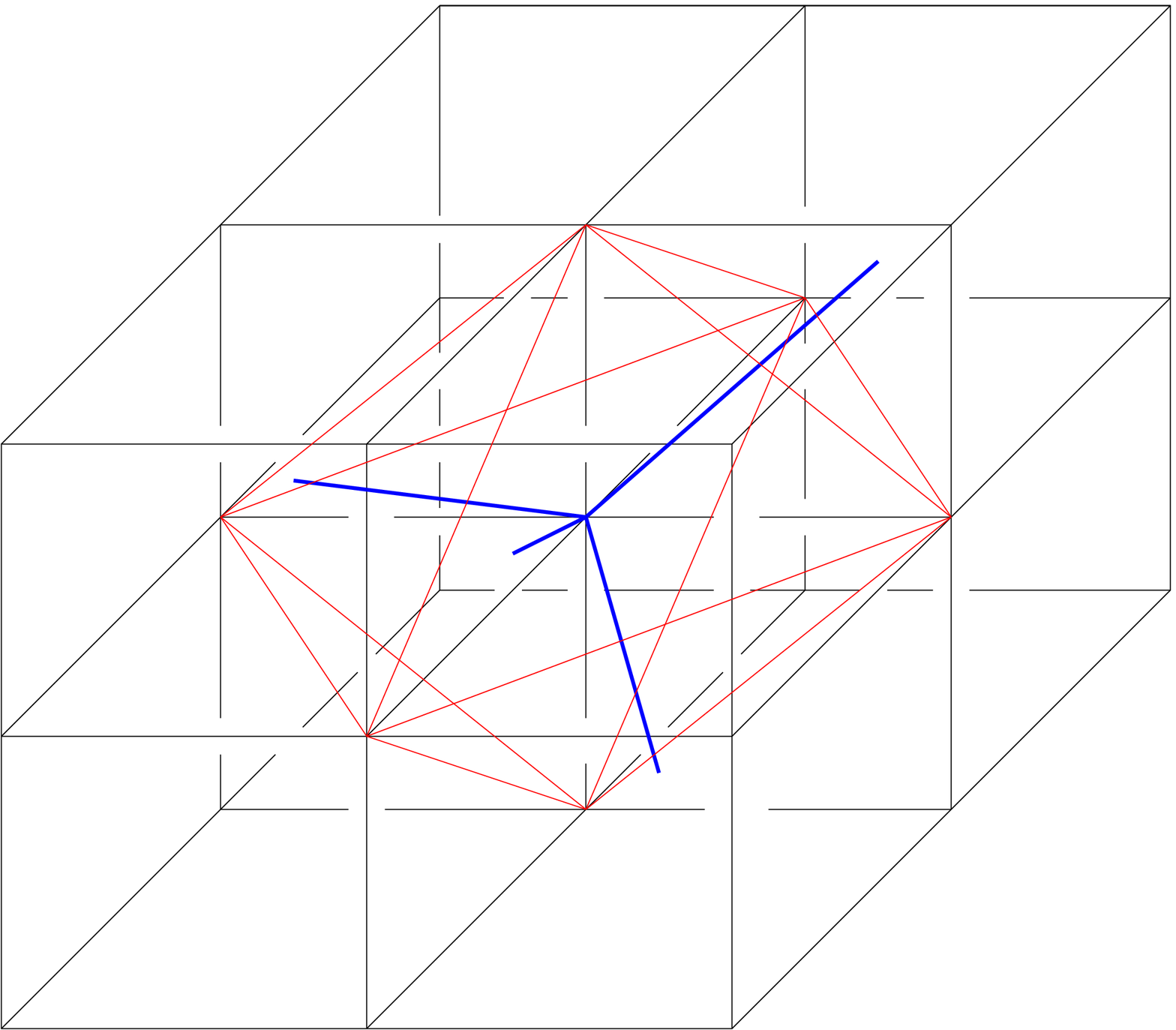}
\caption{Dual diamond cell of type B with only the central four valent
vertex left and keeping only the faces adjacent to the vertex.}
\label{fig18}
\end{center} \end{figure}
\begin{figure}[hbt] 
\begin{center} \includegraphics[scale=0.3]{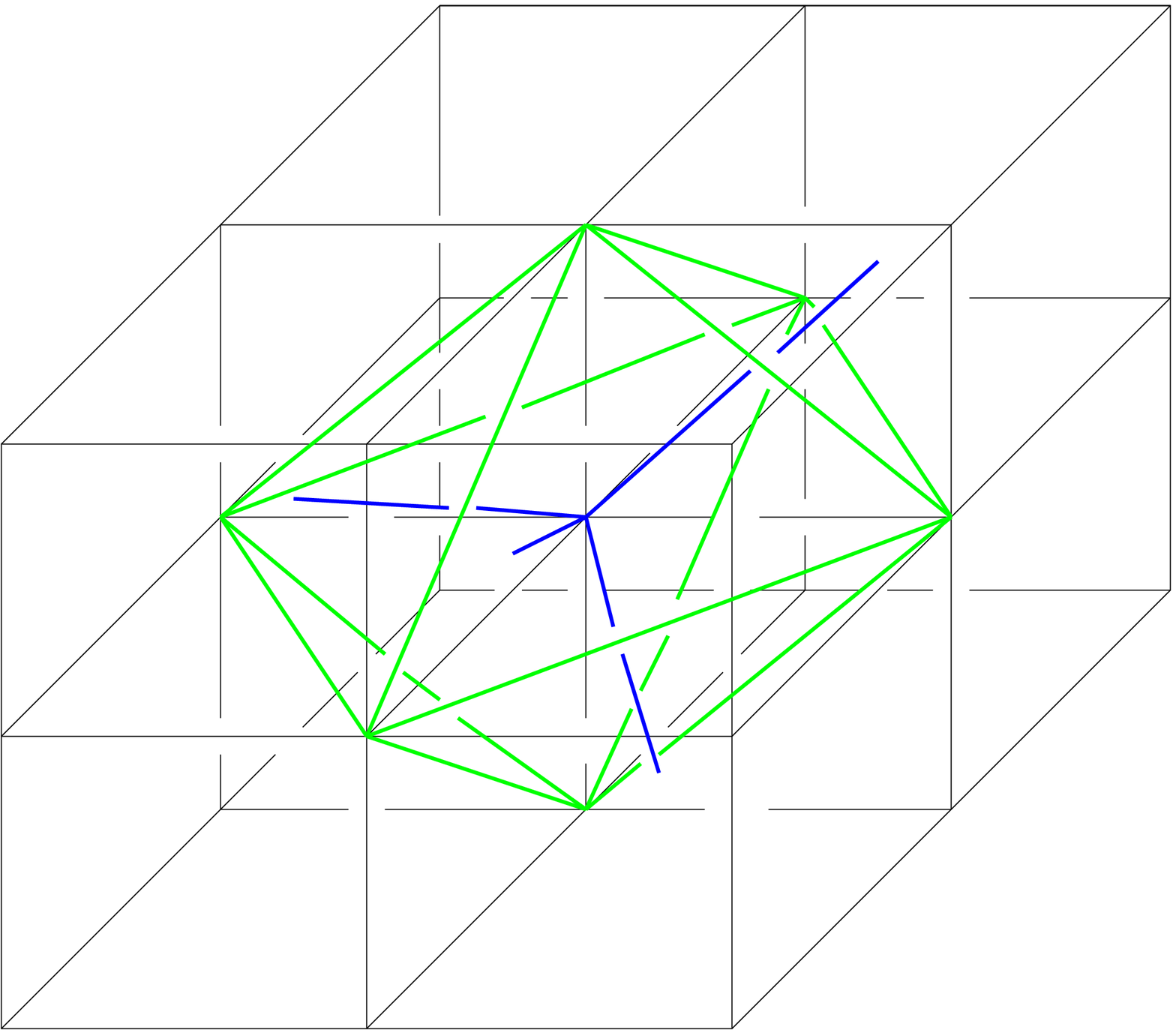}
\caption{Dual diamond cell of type B with only the central four valent
vertex left and keeping only the faces adjacent to the vertex, 
highlighting its octahedronal cell structure.}
\label{fig19}
\end{center} \end{figure}
In order to achieve the desired duality, one has to 
fill in the original cubes again which then triangulate those 
octahedra into eight tetrahedra. This then results in the additional 
verticies and edges that we described and depicted in figures
\ref{fig11} and \ref{fig12}.

\section{Volume Operator Expectation Values for Dual Cell Complex 
Coherent States}
\label{s5} 

In this section we compute the expectation value of the volume operator 
with respect to the dual cell complex coherent states of \cite{10}. 
This section is subdivided into two parts.
In the first we review the definition of the Volume Operator.
In the second, we perform the actual calculation. In 
order to carry it out explicitly, we have to specify the 
graph and the dual cell complex. Here we focus our attention on 
arbitrary graphs with the following properties: 1. All verticies have 
constant valence $n=4,6,8$ and 2. the dual cell complex consists only of 
tetrahedra, cubes and octahedra respectively. Such graphs and dual cell 
complexes exist as we showed explicitly in section \ref{s3}. This is all 
we need for the purposes of this section, more specifics about the graph 
and the complex are not needed. 

\subsection{Review of the Volume Operator}
\label{s4.0}

The classical expression for the volume of a region $R$ of a 
semianalytical three
dimensional manifold $\sigma$ is:
\be \label{4.0.1}
V_R:=\int_R \; d^3x \; \sqrt{\det(q)}=\int_R \; d^3x\; \sqrt{|\det E|}
\ee
where $q_{ab}$ is the three metric. The version of the volume operator 
\cite{8} consistent with the triad quantization \cite{9} that enters the 
quantum dynamics \cite{3} has cylindrically consistent projections 
$\widehat{V}_{R,\gamma}$ given by 
\be \label{4.0.2}
\hat{V}_R=
\sum_{v\in V(\gamma)\cap R}\hat{V}_{\gamma,v}
\ee 
where
\be \label{4.0.3}
\hat{V}_{\gamma,v}=\ell_P^3\; \sqrt{|\frac{1}{8}
\sum_{e_I, e_J, e_K,\\ I\leq J\leq K\leq N|v\in e_I\cap e_J\cap e_K }
\epsilon^{ijk}\epsilon(e_I,e_J,e_K)X^{e_I(v)}_iX^{e_J(v)}_jX^{e_I(v)}_k|}
\ee
Here $N$ denotes the valence of the vertex, $\ell_P^2=\hbar\kappa$ is 
the Planck area, 
$X^{e_I(v)}_i={\rm Tr}([\tau_i h_I]^T \partial/\partial h_I)$ are 
right 
invariant vector on $SU(2)$ acting on the holonomy $h_I:=A(e_I)$
($i \tau_j=\sigma_j$ are the Pauli matrices)  
and $\epsilon(e_I,e_J,e_K)$ is called the orientation function which is 
defined as follows:
\be \label{4.0.4}
\epsilon(e_I,e_J,e_K)=\left\{\begin{array}{lll}
1,\text{iff}\hspace{.05in}\dot{e}_I,\dot{e}_J,\dot{e}_K
\text{are linearly independent at v and positively oriented}\\
-1,\text{iff}\hspace{.05in}\dot{e}_I,\dot{e}_J,\dot{e}_K
\text{are linearly independent at v and negatively oriented}\\
0,\text{iff}\hspace{.05in}\dot{e}_I,\dot{e}_J,\dot{e}_K
\text{are linearly dependent at v}
\end{array}
\right.
\ee
Here we take the convention that the edges at $v$ have been taken with 
outgoing orientation, hence if in $\gamma$ the orientation of an edge 
$e$ adjacent to 
$v$ is actually ingoing, just apply the above expression to 
$\psi'(..,h_e^{-1},...):=\psi(..,h_e,..)$.
 
From (\ref{4.0.2}), we deduce that the volume operator is a sum  
of contributions, one for each vertex. Therefore, in the expectation 
value calculations that follow it will be sufficient to calculate 
the expectation values for each $\hat{V}_{\gamma, v}$ separately and 
then to add the contributions. Notice that each of these contributions 
is of the form $V_{\gamma, v}=\root 4 \of{Q_{\gamma,v}}$ where 
$Q_{\gamma,v}$ is minus the square of the expression appearing between 
the modulus labels $|..|$ in (\ref{3.3}) and therefore is a sixth order 
polynomial in the $SU(2)$ right invariant vector fields.

We will now proceed to calculate the general expression for the 
expectation value of the volume operator for an $n=4,6,8$ valent graph.

\subsection{Expectation Values}
\label{s5.2}

We can actually perform a full $SU(2)$ calculation as follows: \\
The coherent states are explicitly given by \cite{10}
\be \label{5.1}
\psi_{Z,\gamma}=\prod_{e\in E(\gamma)}\; \psi_{Z,e},\;\;
\psi_{Z,e}(A)=\sum_{2j=0}^\infty\; e^{-t j(j+1)/2}\; \chi_j(g_e(Z) 
A(e)^{-1})
\ee
where $t=\ell_P^2/L^2$ and $g_e(Z)$ is given by (\ref{2.35}). 
The volume operator expectation value is given by 
\be \label{5.2}
<V(R)>_{Z,\gamma}=\sum_{v\in V(\gamma)\cap R} 
\;<V_{\gamma,v}>_{Z,\gamma}
\ee
Notice that due to the product form (\ref{5.1}), the expectation value 
$<V_{\gamma,v}>_{Z,\gamma}$ only involves the edges adjacent to $v$. 
Now as we saw in the previous section we have $V_{\gamma,v}=\root 4 
\of{Q_{\gamma, v}}$. By the arguments presented in the introduction, 
the zeroth order in $\hbar$ of $<V_{\gamma,v}>_{Z,\gamma}$ is given by
$\root 4 \of{<Q_{\gamma, v}>_{Z,\gamma}}$. Since $Q_{\gamma,v}$ is a 
polynomial in right invariant vector fields, the results of \cite{10}
reveal that that to zeroth order in $\hbar$ the expectation value of 
any polynomial in the right invariant vector fields $i\ell_P^2 X^j_e$ is 
simply
obtained by replacing it by $E_j(S_e)$ which is given in 
(\ref{2.16a}).

It follows that to zeroth order in 
$\hbar$ we have $<Q_{\gamma,v}>_{Z,\gamma}=[P_{\gamma,v}(E)]^2$ where  
\be \label{5.3}
P_{\gamma,v}(E)=\frac{1}{48}\sum_{e\cap e'\cap e^{\prime\prime}} \;
\epsilon_{e,e',e^{\prime\prime}})\; \epsilon^{jkl}\;
E_j(S_e)\; E_k(S_{e'}) \; E_l(S_{e^{\prime \prime}})
\ee
Notice that for 
sufficiently fine graphs, we can drop the holonomies along the paths 
$\rho_e(x)$ involved in the definition of $E_j(S_e)$ as we approach 
the continuum. It is then clear that the correct
expectation value of the volume operator is reached provided that 
(\ref{5.3}) approximates the volume, as specified by $E^a_j$, of the 
cell of the polyhedronal complex which is bounded by the faces $S_e$
involved in (\ref{5.3}).

To do this, we 
use the fact that for sufficiently fine graphs a polyhedron $P$ in 
$\sigma$ dual to a vertex of the graph lies in the domain of a chart
$Y$ so that $P$ is the image under $Y$ of a standard polyhedron $P_0$
in $\mathbb{R}^3$. Introducing 
\be \label{5.4}
n_a^I(s)=\frac{1}{2}\; \epsilon_{abc} \; \epsilon^{IJK}\;
\frac{\partial Y^b(s)}{\partial s^J}\;
\frac{\partial Y^c(s)}{\partial s^K}\;
\ee
we immediately find with $P=Y(P_0)$ 
\be \label{5.5}
{\rm Vol}(P)=\int_P \; d^3x\; \sqrt{|\det(E)(x)|}
=\int_{P_0} \; d^3s\; \sqrt{|\det(\tilde{E}(s)|}
\ee
where
\be \label{5.6}
\tilde{E}^I_j(s)=E^a_j(Y(s))\; n_a^I(s)
\ee
Now for sufficiently fine graphs (\ref{5.6}) is approximately constant 
over $P_0$ so that 
\be \label{5.7}
{\rm Vol}(P)
\approx  \sqrt{|\det(\tilde{E}(s)|}_{Y(s)=v}\; 
{\rm Vol}_0(P_0)
\ee
where 
\be \label{5.8}
{\rm Vol}_0(P_0)=\int_{P_0} \; d^3s 
\ee
is the volume of the standard polyhedron with respect to the Euclidean
metric on $\mathbb{R}^3$.  

The idea behind this rewriting is that the fluxes $E_j(S_e)$
can be approximated by specific linear combinations of the 
$[\tilde{E}^I_j(s)]_{Y(s)=v}$ 
so that 
a direct comparison between (\ref{5.3}) and (\ref{5.8}) is possible.  
This is because a boundary face $S$ is also the image under $Y$ of a 
standard face $S^0$ in $\mathbb{R}^3$ so that (dropping the holonomies 
along the $\rho_e(x)$ as explained)
\be \label{5.9}
E_j(S)
=\int_S \; \frac{1}{2}\; \epsilon_{abc}\; dx^b\wedge dx^c \;E^a_j(x)
=\int_{S^0} \; \frac{1}{2}\; \epsilon_{IJK}\; ds^J\wedge 
ds^K \;
\tilde{E}^I_j(s)
\approx [\tilde{E}^I_j(s)]_{Y(s)=v} \; {\rm F}^I(S^0)
\ee
where 
\be \label{5.10} 
{\rm F}_I(S^0)
=\int_{S^0} \; \frac{1}{2}\; \epsilon_{IJK}\; ds^J\wedge 
ds^K \;
\ee 
is the $I$ component of the Euclidean flux through $S_0$. Thus, plugging 
(\ref{5.10}) into (\ref{5.3}) we find 
\be \label{5.11}
|P_{\gamma,v}(E)|^{1/2} \approx \sqrt{|\det(\tilde{E}(s))}_{Y(s)=v}\;
{\rm Vol}_0(v)
\ee
where 
\be \label{5.12}
{\rm Vol}_0(v)
=\sqrt{|\frac{1}{48}\sum_{e\cap e'\cap e^{\prime\prime}} 
\;
\epsilon_{e,e',e^{\prime\prime}})\; \epsilon^{IJK}\;
F_I(S^0_e)\; F_J(S^0_{e'}) \; F_K(S^0_{e^{\prime \prime}})|}
\ee

It remains to compare (\ref{5.8}) and (\ref{5.12}).
All of this still holds for general graphs. We now specify to purely 
$n=4,6,8$ valent graphs with the above specified properties in order to 
test the correctness of the expectation value for specific, simple 
situations. Thus we 
know that for each vertex $v$ the faces $S_e$ dual to the edges $e$ 
adjacent to $v$ form the surface a tetrahedron, cube and octahedron 
respectively. Thus we just have to compare (\ref{5.3}) with the volume 
of such platonic bodies as measured by $E^a_j$. 
We will discuss the three cases separately.

\subsubsection{Tetrahedron}
\label{5.2.1}

A standard tetrahedron is the subset
\be \label{5.13}
T_0=\{s\in \mathbb{R}^3:\;0\le s^I\le 1;\;I=1,2,3,\;s^1+s^2+s^3\le 1\}
\ee
It has four boundary triangles given by 
\ba \label{5.14}
t^0_I &=& \{s\in \mathbb{R}^3:\; s^I=0,\; 0\le s^J, s^K\le 
1,\;s^J+s^K\le 1;\;\;\epsilon_{IJK}=1\}
\nonumber\\
t^0_4 &=& \{s\in \mathbb{R}^3:\;0\le s^I\le 1;\;I=1,2,3,\;
s^1+s^2+s^3=1\}
\ea
We easily compute 
\be \label{5.15}
{\rm Vol}_0(T_0)=\frac{1}{6}
\ee
while (remembering that the surfaces carry outward orientation if the 
edges are outgoing from $v$)  
\be \label{5.16}
F_I(t^0_J)=\frac{1}{2} \delta_{IJ},\; F_I(t^0_4)=-\frac{1}{2}
\ee
Let us label the edges adjacent to $v$ by $e_1,..,e_4$ where $e_\alpha$ 
is dual to $Y(t^0_j),\;j=1,2,3,4$. Then
\ba \label{5.17}
{\rm Vol}_0(v) &=&
=\sqrt{|\frac{1}{8}\sum_{1\le j <k<l\le 4} 
\;
\epsilon_{e_j,e_k,e_l})\; \epsilon^{IJK}\;
F_I(t^0_j)\; F_J(t^0_k) \; F_K(t^0_l)
|}
\nonumber\\
&=& \frac{1}{8} \sqrt{|
\epsilon(e_1,e_2,e_3)
-\epsilon(e_1,e_2,e_4)
-\epsilon(e_1,e_3,e_4)
-\epsilon(e_2,e_3,e_3)
|}
\ea
which still depends on the sign factors. Hence the expectation value 
takes values in the range $0,\frac{1}{8},\;
\frac{\sqrt{2}}{8},\;
\frac{\sqrt{3}}{8},\;\frac{1}{4}$
none of which coincides with $\frac{1}{6}$. For the explicit four valent 
graph that we constructed in section (\ref{s3}) each triple among the 
four edges has linearly independent tangents at $v$ and the expectation 
value is given by $\frac{\sqrt{2}}{8} > \frac{1}{6}$ which is too large.

\subsubsection{Cube}
\label{s5.2.2}

A standard cube is the subset
\be \label{5.18}
C_0=\{s\in \mathbb{R}^3:\;0\le s^I\le 1;\;I=1,2,3\}
\ee
It has six boundary squares given by 
\ba \label{5.19}
s^0_{I +} &=& \{s\in \mathbb{R}^3:\; s^I=1,\; 0\le s^J, s^K\le 
1;\;\;\epsilon_{IJK}=1\}
\nonumber\\
s^0_{I -} &=& \{s\in \mathbb{R}^3:\; s^I=0,\; 0\le s^J, s^K\le 
1;\;\;\epsilon_{IJK}=1\}
\ea
We easily compute 
\be \label{5.20}
{\rm Vol}_0(T_0)=1
\ee
while (remembering that the surfaces carry outward orientation if the 
edges are outgoing from $v$)  
\be \label{5.21}
F_I(s^0_{J \sigma})=\sigma \delta_{IJ}
\ee
with $\sigma=\pm$. 

Let us label the edge dual to $Y(s^0_{I \sigma})$ by $e_{I\sigma}$. Then 
the expectation value becomes 
\be \label{5.22}
{\rm Vol}(v)=
\sqrt{|
\frac{1}{48} \sum_{I,J,K;\sigma_1,\sigma_2,\sigma_3} \;
\epsilon(
e_{I\sigma 1},e_{J\sigma 2},   
e_{K\sigma 3})\; \sigma_1\sigma_2\sigma_3 \; \epsilon_{IJK}
|} \ee
which again depends on the precise embedding of the graph. For an actual 
cubical graph constructed in section \ref{s3}, the edges $e_{I +}, e_{I 
-}$ are analytic continuations of 
each other so that the orientation factor vanishes if two or more edges 
carry the same direction label $I$. Otherwise there are more 
contributions. Which orientation factors are allowed has been analyzed in 
detail in \cite{17}. In the case of the actual cubical graph we have 
$\epsilon(
e_{I\sigma 1},e_{J\sigma 2},   
e_{K\sigma 3})=\sigma_1 \sigma_2 \sigma_3 \epsilon_{IJK}$ so that 
(\ref{5.22}) becomes
\be \label{5.23}    
{\rm Vol}(v)=
\sqrt{|
\frac{1}{6} \sum_{I,J,K} \; \epsilon_{IJK}^2|}=1
\ee
which coincides with (\ref{5.20}).

\subsubsection{Octahedron}
\label{s5.2.3}

A standard octahedron is the subset
\be \label{5.24}
O_0=\{s\in \mathbb{R}^3:\; |s^3|\le \frac{1}{2},\; |s^1|,|s^2| \le
\frac{1}{2} -|s^3|\}
\ee
It has eight boundary triangles given by 
\be \label{5.25}
t^0_{I \sigma \sigma'} = \{s\in \mathbb{R}^2:\; 
0\le \sigma' s^3 \le \frac{1}{2},\;s^I=\sigma(\frac{1}{2}-|s^3|),\;
|s^J|\le \frac{1}{2} -|s^3|\}
\ee
where $I,J=1,2;\;I\not=J;\;\sigma,\sigma_3=\pm$. 

We easily compute 
\be \label{5.26}
{\rm Vol}_0(O_0)=\frac{1}{3}
\ee
while (remembering that the surfaces carry outward orientation if the 
edges are outgoing from $v$)  
\be \label{5.27}
F_I(t^0_{J \sigma \sigma'})=\frac{1}{4}[\sigma \delta_{IJ}+\sigma' 
\delta_{I3}]
\ee

Labelling the edge dual to $Y(t^0_{I \sigma \sigma_3})$ by $e_{I \sigma 
\sigma_3}$ we find for the expectation value 
\be \label{5.28}
{\rm Vol}_0(v) = \sqrt{|
\frac{1}{48\cdot 64} \sum_{{I_1,I_2,I_3=1,2} \atop
{\sigma_1,\sigma_2,\sigma_3,
\sigma'_1,\sigma'_2,\sigma'_3=\pm}}\;
\epsilon(
e_{I_1 \sigma_1 \sigma_1'},
e_{I_2 \sigma_2 \sigma_2'},
e_{I_3 \sigma_3 \sigma_3'})\;
[\sigma_1 \sigma_2 \sigma_3' \epsilon^{I_1 I_2}
+\sigma_1 \sigma_2' \sigma_3' \epsilon^{I_3 I_1}
+\sigma_1' \sigma_2 \sigma_3 \epsilon^{I_2 I_3}]
|}
\ee
where $\epsilon^{IJ}$ is the alternating 
symbol for $I,J=1,2$ with $\epsilon^{12}=1$. Expression (\ref{5.28})
is already very complicated to analyse for the most general 
edge configuration and again we refer to \cite{17} for a comprehensive 
discussion. However, for the case of the 
graphs constructed in section \ref{s3} the situation becomes simple 
enough. Namely in this case the eight edges $e_{I \sigma \sigma'}$ have 
the property that $e_{I, sigma, \sigma'}$ and $e_{I, -\sigma, -\sigma'}$ 
are analytic 
continuations of each other. This implies that $\dot{e}_{I, \sigma, 
\sigma'}(0)=\sigma' \dot{e}_{I, \sigma\sigma',+}(0)$ where $e_{I \sigma 
\sigma'}(0)=v$ is the common starting point of all edges. Since 
$\epsilon(e,e',e^{\prime\prime})={\rm sgn}(\det(\dot{e}(0),\dot{e}'(0),
\dot{e}^{\prime\prime}(0)))$ is completely skew in 
$e,e',e^{\prime\prime}$, in this case we can simplify (\ref{5.28}) 
to
\ba \label{5.29}
{\rm Vol}_0(v) &=& \sqrt{|
\frac{1}{48\cdot 64} \sum_{{I_1,I_2,I_3=1,2}\atop
{\sigma_1,\sigma_2,\sigma_3,\sigma'_1,\sigma'_2,\sigma'_3=\pm}}\;
\epsilon(
e_{I_1, \sigma_1 \sigma_1',+},
e_{I_2, \sigma_2 \sigma_2',+},
e_{I_3,\sigma_3 \sigma_3',+})
}
\nonumber\\
&& \overline{\times\;
[\sigma_1\sigma_1' \sigma_2 \sigma_2' \epsilon^{I_1 I_2}
+\sigma_1\sigma_1' \sigma_3 \sigma_3' \epsilon^{I_3 I_1}
+\sigma_2\sigma_2' \sigma_3 \sigma_3' \epsilon^{I_2 I_3}]
|}
\ea
Since (\ref{5.29}) only depends on $\tilde{\sigma}_I=\sigma_I \sigma_I'$,
after proper change of summation variables, (\ref{5.29}) turns into 
\be \label{5.29a}
{\rm Vol}_0(v) = \sqrt{|
\frac{1}{48\cdot 8} \sum_{{I_1,I_2,I_3=1,2}\\
{\sigma_1,\sigma_2,\sigma_3=\pm}}\;
\epsilon(
e_{I_1, \sigma_1,+},
e_{I_2, \sigma_2,+},
e_{I_3, \sigma_3,+})\;
[\sigma_1 \sigma_2 \epsilon^{I_1 I_2}
+\sigma_1 \sigma_3 \epsilon^{I_3 I_1}
+\sigma_2 \sigma_3 \epsilon^{I_2 I_3}]
|}
\ee
Using that $\epsilon(
e_{I_1, \sigma_1,+},
e_{I_2, \sigma_2,+},
e_{I_3, \sigma_3,+})$ and $\sigma_1 \sigma_2 \epsilon^{I_1 I_2}$ are 
both antisymmetric under the simultaneous exchange $(\sigma_1 
I_1) \leftrightarrow (\sigma_2 I_2)$ etc. we may further simplify 
(\ref{5.29a}) to 
\ba \label{5.30}
{\rm Vol}_0(v) &=& \sqrt{|
\frac{1}{48\cdot 4} \sum_{\sigma_1,\sigma_2,\sigma_3=\pm}\;
[\sum_{I_3} \;\sigma_1 \sigma_2\;
\epsilon(e_{1, \sigma_1,+},e_{2, \sigma_2,+},e_{I_3, \sigma_3,+})
+\sum_{I_1} \;\sigma_2 \sigma_3\;
\epsilon(e_{I_1, \sigma_1,+},e_{1, \sigma_2,+},e_{2, \sigma_3,+})
}
\nonumber\\
&& \overline{+\sum_{I_2} \;\sigma_3 \sigma_1\;
\epsilon(e_{1, \sigma_1,+},e_{I_2, \sigma_2,+},e_{2, \sigma_3,+})
]
|}
\ea
Carrying out the respective sums over $I_1,I_2,I_3$ and using that 
$\epsilon(e,e',e^{\prime\prime})$ is completely skew we can bring all
orientation factors into one of the two  
standard forms 
$\epsilon(e_{1,\sigma_1,+},e_{1,\sigma_2,+},e_{2,\sigma_3,+})$
and
$\epsilon(e_{2,\sigma_1,+},e_{2,\sigma_2,+},e_{1,\sigma_3,+})$
respectively. After proper relabelling of the $\sigma_I$ we find that
\be \label{5.31}
{\rm Vol}_0(v) = \sqrt{|
\frac{1}{16\cdot 4} \sum_{\sigma_1,\sigma_2,\sigma_3=\pm}\; \sigma_3\;
[\sigma_2\;
\epsilon(e_{1, \sigma_1,+},e_{1, \sigma_2,+},e_{2, \sigma_3,+})
+\sigma_1\;
\epsilon(e_{2, \sigma_1,+},e_{2, \sigma_2,+},e_{1, \sigma_3,+})
]
|}
\ee
Since $\epsilon(e_{I, \sigma_1,+},e_{I, \sigma_2,+},e_{J, \sigma_3,+})$
is skew in $\sigma_1,\sigma_2$ the sum over $\sigma_2$ 
collapses to the term $\sigma_2=-\sigma_1$ and (\ref{5.31}) becomes
\ba \label{5.32}
{\rm Vol}_0(v) 
&=& \sqrt{|
\frac{1}{16\cdot 4} \sum_{\sigma_1,\sigma_3=\pm}\; 
\sigma_3\;\sigma_1\;
[-
\epsilon(e_{1, \sigma_1,+},e_{1, -\sigma_1,+},e_{2, \sigma_3,+})
+
\epsilon(e_{2, \sigma_1,+},e_{2,-\sigma_1,+},e_{1, \sigma_3,+})
]
|}
\nonumber\\
&=& \sqrt{|
\frac{1}{16\cdot 2} \sum_{\sigma_3=\pm}\; 
\sigma_3\;
[-\epsilon(e_{1,+,+},e_{1,-,+},e_{2, \sigma_3,+})
+
\epsilon(e_{2,+,+},e_{2,+,+},e_{1, \sigma_3,+})
]
|}
\ea
Finally, using 
$\epsilon(e_{I,+,+},e_{I,-,+},e_{J, \sigma_3,+})=
\sigma_3\; \epsilon(e_{I,+,+},e_{I,-,+},e_{J,+,+})$ 
and \\
$\epsilon(e_{1,+,+},e_{1,-,+},e_{2,-,+})=
\epsilon(e_{2,+,+},e_{2,-,+},e_{2,+,+})=1$ 
we find 
\be \label{5.33}
{\rm Vol}_0(v)=\frac{1}{2\sqrt{2}}
\ee
which does not agree with (\ref{5.26}). 

\subsection{Discussion}
\label{s5.3}

Interestingly, for both valence 
$n=4$ or $n=8$ the expectation value is larger than the expected value 
with the same ratio $3/(2\sqrt{2})$. 
In general, for generic edge 
configurations and for higher and higher valence the 
expectation value will probably also be larger in ratio than the 
expected 
volume.   
This is because for a vertex of valence $n$ the number of ordered 
triples of edges contributing to the expectation value is given by 
${n\choose 3}$ and for appropriate choice of the orientation factors,
these terms all contribute with the same sign. Such a choice is always 
possible up to topological 
obstructions discussed to some extent
in \cite{17}. For large $n$ the polyhedron dual to the vertex will 
approach more and more a sphere triangulated into $n$ polygonal faces of 
typical 
unit area $4\pi/n$. Hence we expect the leading $n$ 
behavior of the expectation value to be 
given by $\sqrt{\frac{1}{8}\; n^3/6\; 
(4\pi/n)^3}=\sqrt{8\pi^3/6}=4\pi/3\sqrt{3\pi/4}$ while 
the expected volume should approach $4\pi/3$.    

Surely, we have not shown that for graph topologies different from a 
cubical one the expectation value of the volume operator with 
respect to the dual cell complex coherent states cannot be matched with 
the classical volume value. This is because one can allow degenerate 
triples which decrease the volume expectation value. However, the 
discussion reveals that the question for which graphs the expectation 
value comes out correctly is far from trivial and even for natural 
choices the only admissible graph topology is the cubical one.

Notice that the expectation value is insensitive to the embedding of the 
graph relative to the dual cell complex as long as the graph is dual to 
it. For non dual embeddings or graph topologies which do not match the 
cell complex topology at all, the expectation value will be completely 
off the correct value. This demonstrates that the cut -- off graph must 
lie within a certain class which is adapted to the cell complex.

\section{Summary and Conclusions}
\label{s6}

Together with the analysis in \cite{30b} we have shown that the only 
known states of LQG which are semiclassical 
for the volume operator must be based on cubic cut -- off graphs.
This looks surprising at first but can maybe be understood intuitively 
as follows:
 
The volume operator is a derived operator and arises from the 
known representation of the flux operator on the Hilbert space. 
The derivation involves a regularization step which involves cubes 
surrounding the verticies of the graph in question on whose faces 
the fluxes are located. In order to take the limit in which the cubes 
shrink to the verticies and in order to make the result independent of 
the relative orientation between cubes and graphs, an averaging 
procedure must be applied. Hence one might be tempted to say that the 
fact that cubical topology is singled out rests on the cubical 
regularization. 

However, this is not the case. Namely, cylindrical 
consistency and background 
independence alone already fix the cylindrical projections of the 
volume operator up to a global constant as proved explicitly in 
\cite{7,8}. The constant depends on the averaging procedure chosen 
and on whether one uses tetrahedra rather than cubes in the 
regularization. However, 
consistency between volume and flux quantization fixes 
that factor \cite{9} and rules out the operator \cite{7}. That is to 
say, there is no freedom left in defining the volume operator and 
therefore the details of the regularization do not matter, it is a 
regularization independent result. Hence, the preference for cubic 
graphs in the semiclassical analysis must have a different origin.

To see what it is, notice that the volume operator at a vertex involves 
a sum over ordered triples of edges adjacent to the vertex of which  
only the those with 
linearly independent tangents contribute. If the vertex has valence $n$
then typically there are ${n \choose 3}$ contributions \cite{17}. 
They all contribute with equal weight (up to sign) which is the unique
factor determined in \cite{9}. That constant is such that each triple 
contributes as if (the tangents of) a triple of edges spans a 
corresponding 
parallelepiped. However, it is clear that generally far less than 
${n \choose 3}$ parallelepiped are sufficient to triangulate a 
(dual) neighborhood of the vertex and thus it is not surprising that 
large
valence cut -- off graphs will not give rise to good semiclassical 
states. On the other hand, unless the graph is cubic, even at low 
$n=4$ the parallelepiped volume contribution per triple is too high
for the triangulation of a tetrahedron. We have seen both effects at 
work in the previous section.  

This result has two implications: Either one is able to find new types 
of states which are not constructed by the complexifier method or by 
different complexifiers than the ones employed so far such that the 
correct semiclassical behavior is recovered also for graphs of 
different than cubic topology. Or, if that turns out to be impossible,
one should accept this result and conclude that, in order that the 
boundary Hilbert space of spin foam models has a semiclassical sector,
one should generalize them to more general than simplicial 
triangulations of the four manifold as advocated in \cite{30,30a}.\\   
\\
\\
{\large\bf Acknowledgments}\\
\\C.F. is grateful to the Perimeter Institute for Theoretical Physics for hospitality and ﬁnancial sup-
port where parts of the present work were carried out.
Research performed at Perimeter
Institute for Theoretical Physics is supported in part by the
Government of Canada through NSERC and by the Province of Ontario
through MRI.

\end{document}